\documentclass[preprint,aps]{revtex4}

\usepackage{graphicx}
\usepackage{dcolumn}
\usepackage{bm}

\usepackage{amsmath}
\usepackage{graphicx}
\usepackage{amssymb}
\usepackage{mathrsfs}
\usepackage{bbm}

\renewcommand{\vec}[1]{{\boldsymbol{#1}}}
\newcommand{\la}{\label}

\newcommand{\be}{\begin{equation}}
\newcommand{\ee}{\end{equation}}
\newcommand{\ba}{\begin{eqnarray}}
\newcommand{\ea}{\end{eqnarray}}
\newcommand{\bastar}{\begin{eqnarray*}}
\newcommand{\eastar}{\end{eqnarray*}}

\newcommand{\ce}{{\mathbbmss e}}
\begin{document}

\begin{titlepage}

\title{{\normalsize \bf SPIN-CHARGE SEPARATION,  CONFORMAL COVARIANCE \\
AND THE SU(2) YANG-MILLS THEORY}}
\vskip 2.0cm
\author{Ludvig Faddeev}
\email{Faddeev@pdmi.ras.ru}
\affiliation{S. Petersburg Branch of Steklov Mathematical Institute,
Russian Academy of Sciences, Fontanka 27, St. Petersburg, Russia
}
\author{Antti J. Niemi}
\email{Antti.Niemi@teorfys.uu.se}
\homepage{http://www.teorfys.uu.se/people/antti}
\affiliation{Department of Theoretical Physics,
Uppsala University,
P.O. Box 803, S-75108, Uppsala, Sweden}
\affiliation{
Laboratoire de Mathematiques et Physique Theorique
CNRS UMR 6083, Universit\'e de Tours, 
Parc de Grandmont, F37200, Tours, France}
\affiliation{ Chern Institute of Mathematics, 
Tianjin 300071, P.R. China } 
\date{\today}
 
\vskip 2.0cm
\begin{abstract}
In the low energy domain of four-dimensional SU(2) 
Yang-Mills theory the spin and the charge of the gauge field 
can become separated from each other. The ensuing field variables
describe the interacting dynamics between a version 
of the $O(3)$ nonlinear $\sigma$-model and a nonlinear Grassmannian 
$\sigma$-model, both of which may support closed knotted strings 
as stable solitons. Lorentz transformations act 
projectively in the $O(3)$ model which breaks 
global internal rotation symmetry and removes massless Goldstone 
bosons from the particle spectrum. The entire Yang-Mills Lagrangian 
can be recast into a generally covariant form with a conformally 
flat metric tensor. The result contains the 
Einstein-Hilbert Lagrangian together with 
a nonvanishing cosmological constant, and insinuates the presence of
a novel dimensionfull parameter in the Yang-Mills theory.
\end{abstract}
\maketitle
\end{titlepage}



\section{Introduction}
Apparently the necessity of a mass 
gap in a pure Yang-Mills theory and the nature of its 
particle spectrum were originally 
posed as problems by Wolfgang Pauli during 
a 1954 Princeton seminar by C.N. Yang \cite{Yang}. 
However, despite over 50 years of effords the physical 
particle content of a pure, interacting four dimensional 
Yang-Mills theory remains a mystery.
In particular, the explanation of color confinement
is a theoretical challenge \cite{clay}. 
Dimensional transmutation with its intimate relationship 
to the high energy asymptotic freedom does open 
a door for a dimensionfull parameter to enter.
But since the high energy limit of the Yang-Mills theory
describes asymptotically free, massless gauge particles 
the precise relationship between
this dimensionfull parameter and the mass gap 
which secures confinement remains 
unclear. 

During the last ten years \cite{omat1}-\cite{omat3}
we have investigated the possibility that the low energy
spectrum of pure Yang-Mills theory could comprise of
closed and knotted strings as stable solitons. 
This proposal is very natural from the point of view of 
QCD phenomenology. If quarks are indeed confined by 
stringlike collective excitations of the gauge field, 
in the absence of quarks these strings should close 
on themselves into stable and in general knotted and
linked configurations. In \cite{omat3} we proposed 
how the properties of such closed stringlike solitons 
could be related to the spectrum of the 
Yang-Mills theory. In particular, 
we suggested that in the case of a $SU(2)$ gauge theory 
the effective low energy Lagrangian should relate to the 
following version of the $O(3)$ nonlinear $\sigma$-model,
originally proposed by one of us \cite{fadpap}
\begin{equation}
{\mathcal L}_{eff} \ = \ 
\frac{\mathrm m^2}{2} (\partial_a \vec s)^2 + \frac{1}{4}
(\vec s \cdot \partial_a \vec s
\times \partial_b \vec s)^2 + V(\vec s)
\la{0fha}
\end{equation}
Here $\vec s$ is a three-component unit vector, 
and $\mathrm m$ is a parameter with
the dimensions of mass. The last term is a potential term. It 
breaks the global $O(3)$ symmetry which is present in the 
first two terms, removing the two massless 
Goldstone bosons from the spectrum. Both numerical simulations 
\cite{jarmo} and formal mathematical 
arguments \cite{wu} have confirmed our proposal \cite{nature}
that (\ref{0fha}) does indeed support closed knotted strings as
stable solitons.

Several approaches have been suggested, how to derive (\ref{0fha})
directly from the Yang-Mills theory. 
These derivations are commonly based
on the following, {\it incomplete} Lie-algebra
expansion of the $SU(2)$ gauge field \cite{omat1}
\begin{equation}
\vec A_a = C_a \vec s + \partial_a \vec s \times
\vec s + \rho \partial_a \vec s + \sigma \partial_a \vec s
\times \vec s
\la{cho}
\end{equation} 
where $\rho$ and $\sigma$ are real scalars. 
The first two terms in (\ref{cho}) were 
originally introduced by Duan and Ge \cite{ge}, 
and subsequently by Cho \cite{chos}, to describe the properties 
of Wu-Yang monopoles. Indeed, this is a very natural decomposition 
of the gauge field in terms of the $SU(2)$ Lie algebra. 
When (\ref{cho}) is substituted to the Yang-Mills Lagrangian, one 
finds that the structure (\ref{0fha}) 
emerges after one-loop radiative corrections are taken into 
account \cite{edw}. 

The apparent lack of off-shell completeness in (\ref{cho}) 
prompted us in \cite{omat2}
to propose an alternative decomposition of
the gauge field. This new decomposition 
is off-shell {\it complete}, 
and (\ref{cho}) assumes a role in ensuring its 
manifest gauge covariance. Our new decomposition 
leads to many unexpected consequences, some of which we shall 
reveal in the present article. In particular, it admits 
a very intriguing physical interpretation: 
The decomposition \cite{omat2} is intimately related to 
the slave-boson decomposition  \cite{and1} that has been
introduced independently in the context of strongly correlated
electron systems as an alternative to 
Cooper pairing and Higgs effect \cite{and2}. 
As a consequence our new decomposition suggests that the 
separation between the spin and the charge could
be a general phenomenon that can be
exhibited by a large variety of quantum fields
\cite{walet}, \cite{cherno}. 
 
A common feature of these spin-charge decompositions is that
they all seem to involve 
a real-valued scalar field. A nonvanishing ground state 
expectation value for this scalar field describes a condensate. 
This corresponds to the density of the material environment, and 
the presence of a nontrivial condensate 
is a necessary condition for the spin-charge decomposition 
to occur. The condensate can also yield
an alternative to the conventional Higgs effect \cite{and2}.

Here we find that in the 
case of $SU(2)$ Yang-Mills theory the real-valued scalar field admits
another interpretation. It is
the conformal scale of a metric tensor that
describes a conformally flat space-time. In terms of
the spin-charge separated variables, the Yang-Mills 
Lagrangian then contains both the Einstein-Hilbert Lagrangian
and a cosmological constant. Since conformal 
flatness is equivalent to the vanishing of the traceless 
Weyl (conformal) tensor $W_{\mu\nu\rho\sigma}$, the gravitational 
contribution to the Yang-Mills Lagrangian 
can be interpreted as the $\gamma \to 
\infty$ limit of the higher derivative gravitational
Lagrangian
\begin{equation}
{\mathcal L}_{EH}^{} \ = \ \frac{1}{\kappa} 
\sqrt g \, R \ - \ \Lambda \cdot \sqrt{g} \ + \
\gamma \cdot W^{2}_{\mu\nu\rho\sigma}
\la{ehacti}
\end{equation}
It has been shown \cite{stelle} that the Lagrangian (\ref{ehacti}) 
is part of a {\it renormalizable} higher derivative
quantum theory of gravity.
In particular, the one-loop $\beta$-function for 
$\gamma$ does indeed send this coupling to infinity in the
short distance limit. This enforces asymptotically the condition
\begin{equation}
W_{\mu\nu\rho\sigma} \ \sim \ 0
\la{weylc}
\end{equation}
Hence at short distances the space-time becomes asymptotically
(locally) conformally flat. We also note that the 
presence of the higher derivative
Weyl tensor contribution in (\ref{ehacti})
gives rise to a linearly increasing component
in the large distance interactions. Furthermore, our 
reformulation of the Yang-Mills theory
in a manner that contains the Einstein-Hilbert and cosmological
constant terms has the additional
peculiar consequence that a novel  
dimensionfull parameter enters the Yang-Mills Lagrangian. 

\vskip 0.3cm

We start the present article by
defining our notations. We then proceed to describe the
new decomposition \cite{omat2} of the four 
dimensional $SU(2)$ gauge field. 
This decomposition singles out the Cartan direction of the $SU(2)$
Lie algebra. This leaves us with a geometric structure that involves
the Grassmannian manifold $G(4,2)$ of two dimensional planes that 
are embedded in a four dimensional space. 
The Grassmannian geometry guides our subsequent decomposition 
of the gauge field into its spin and charge constituents. In particular,
it leads us to a $O(3)$ symmetric order 
parameter, a three component internal space unit 
vector field $\vec n$ akin the vector field $\vec s$ in (\ref{0fha}).

The Grassmannian structure also introduces an internal, compact 
$U(1)$ gauge symmetry. In the context of lattice gauge theories \cite{hay}
a compact $U(1)$ gauge theory is known to display
a first-order phase transition between a strong coupling 
phase and a weak coupling phase. The strong coupling phase
exhibits confinement, which is absent in the weak coupling phase.   

It is quite natural to expect that in an abelian theory
the gauge coupling decreases when the distance increases.
Thus the presence of the internal compact $U(1)$ gauge 
structure may explain why in short distance Yang-Mills theory 
the spin and the charge can become strongly confined into 
asymptotically free and pointlike gauge particles, and why 
the decomposition into independent spin and charge carriers can 
only occur in the weakly coupled long-distance domain of the compact $U(1)$
theory.

The internal $U(1)$ gauge structure
leads to a projective realization of the Lorentz transformations in the
$O(3)$ unit vector $\vec n$. The ensuing one-cocycle breaks the 
global $O(3)$ symmetry which is displayed by the
first two terms in the corresponding Lagrangian (\ref{0fha}),
even in the absence of an explicit symmetry breaking term such as the third
term in (\ref{0fha}): Since the
ground state of the theory can not violate Lorentz invariance, 
the ground state direction of the vector $\vec n$ in the internal space 
becomes uniquely fixed. As a consequence the requirement that 
the ground state is Lorentz invariant, removes the two massless Goldstone 
bosons that are otherwise associated with the breaking
of the global $O(3)$ symmetry in the dynamics of $\vec n$.

We then proceed to inspect the detailed structure of the
Yang-Mills Lagrangian in 
terms of the separate spin and charge variables. 
In particular, we explain how the Lagrangian (\ref{0fha}) for the
internal vector $\vec n$ is embedded in the tree-level 
Yang-Mills Lagrangian, even before any radiative corrections 
are taken into account. Since (\ref{0fha}) supports closed 
knotted strings as stable solitons \cite{nature},
\cite{jarmo}, \cite{wu}, this endorses our proposal that knotted and closed 
stringlike solitons are indeed the natural candidates for describing 
the interacting spectrum of the pure $SU(2)$ Yang-Mills 
theory \cite{omat3}.

It turns out that the functional form
(\ref{0fha}) relates both to the spin and the charge
degree of freedom. This suggests the presence of some 
kind of dual structure between the spin and charge variables 
in the Yang-Mills theory.
Moreover, since the spin conventionally relates to magnetism 
while the charge relates to electricity, this could be the
sought-after electric-magnetic duality of the Yang-Mills theory.

We then argue that our decomposition is independent 
of the way how we choose the Cartan direction in the $SU(2)$ 
Lie algebra. The gauge covariance becomes manifest when
we introduce the structure of (\ref{cho}) in
our decomposed gauge field.

We show that in terms of the spin and the charge variables
the entire Yang-Mills Lagrangian admits a manifestly generally 
covariant form with a conformally flat metric tensor. 
Thus the $SU(2)$ Yang-Mills theory describes
the interactions between (\ref{0fha})  and the $G(4,2)$ Grassmannian
nonlinear $\sigma$-model in a conformally flat space-time.
Both the Einstein-Hilbert Lagrangian and a cosmological
constant term are present. This also introduces a
dimensionfull parameter in the Yang-Mills theory, that
becomes visible only when it is realized in terms of the
spin-charge separated variables. 

Our result suggests the tantalizing
possibility that long distance Einstein gravity metamorphoses
into a renormalizable Yang-Mills theory at short distances.

Finally, we analyze the finite energy content of the
spin-charge separated, static Yang-Mills theory using a Hamiltonian
formulation. We find that closed knotted strings can also
be supported by the $G(4,2)$ nonlinear $\sigma$-model, in a manner
that involves the structure of (\ref{0fha}). Furthermore,
we propose that for finite energy configurations 
the space-time $\mathbb R^4$ becomes compactified
into $\mathbb S^3\times \mathbb R^1$.

We conclude by presenting some 
physical interpretations of our results, together
with suggestions on the directions for future
research.


\vfill\eject



\section{Some notation}



We consider a $SU(2)$ Yang-Mills gauge field
$ A = A_a^i \sigma^i dx^a $ in $\mathbb R^4$ ($a,b, 
... = 1,2,3,4$)
with Euclidean signature. The Lie algebra generators coincide
with the standard Pauli matrices $\sigma^i \ (i,j, ... =1,2,3)$
and we employ the complex
combinations
\begin{equation}
\sigma^{\pm} = \frac{1}{2}( \sigma^1 \pm i \sigma^2)
\la{pauli}
\end{equation}
For the gauge field this gives
\begin{equation}
A =  A_a^i \sigma^i dx^a = A_a \sigma^3 dx^a + 
X^+_a \sigma^- dx^a + X^-_a \sigma^+ dx^a
\la{pauli2}
\end{equation}
where
\[
X_a^\pm = A^1_a \pm i A^2_a
\]
The finite gauge transformation is
\begin{equation}
A \ \to \ \mathit g A \mathit g^{-1} + 2i \mathit g d \mathit
g^{-1}
\la{2fin}
\end{equation}
Note that in our notation the Yang-Mills coupling
constant appears as a factor in front of the Lagrangian.

For an infinitesimal group element
\[
\mathit g = \exp\{ \frac{i}{2} \vec \epsilon \cdot \vec \sigma \}
\approx 1 + \frac{i}{2} \hat \epsilon + {\cal O}(\epsilon^2)
\]
the gauge transformation takes the form
\[
\delta_{\hat \epsilon} A_a^i \ = \ 
\delta^{ij} \partial_a \epsilon^j
+ \varepsilon^{ijk} A_a^k \epsilon^k
\]
When the gauge transformation is in the direction of the
Cartan subgroup $U^{}_C(1) \in SU(2)$
\begin{equation}
\mathit g \ \sim \ \mathit h = e^{ \frac{i}{2} \omega \sigma^3} 
\in U^{}_C(1)
\la{h}
\end{equation}
the component $A_a^3 \sim A_a$ transforms as 
a $U^{}_{C}(1)$ gauge field
\begin{equation}
\delta_{\mathit h} A_{a} \ = \ \partial_a \omega
\la{abh1}
\end{equation}
For the off-diagonal $X^\pm_a$ we get
\begin{equation}
\delta_{\mathit h} X_a^\pm \ = \ e^{\mp i \omega } X_a^\pm
\la{abh2}
\end{equation}
Consequently, when we only consider gauge transformations in
the direction of the Cartan subgroup, we can interpret 
the full $SU(2)$ 
gauge field as a charged $U^{}_C(1)$ 
vector multiplet
\[
A_a^i \ \to \
(A^{}_a, X_a^\pm)
\]

Eventually we shall argue that even though we here introduce a 
particular (global) identification of the Cartan $U^{}_C(1)$ in terms
of the Pauli matrices, our results are independent
of this particular choice.  However, for the clarity of
presentation we shall momentarily proceed with
this choice of the Cartan direction in the $SU(2)$ Lie-algebra.

Finally, the Yang-Mills field strength tensor is
\[
F_{ab}^i= \partial_a   A_b^i - 
\partial_b A_a^i + \epsilon^{ijk} A_a^j A_b^k
\]
In terms of the charged $U^{}_C(1)$ vector multiplet 
it decomposes according to
\begin{equation}
F^3_{ab} = \partial_a A_b - \partial_b A_a
+ \frac{i}{2} (X_a^+ X_b^- - X_b^+ X_a^-)
\ = \ F_{ab} + P_{ab}
\la{F3}
\end{equation}
\begin{equation}
F_{ab}^\pm = F^1_{ab} \pm i F^2_{ab} =
(\partial_a\pm i A_a) X_b^\pm
- (\partial_b \pm i A_b)X_a^\pm \ = \ 
\mathrm D_{A\, a}^\pm \, X_b^\pm - \mathrm D_{A\, b}^\pm \, X_a^\pm
\la{F+-}
\end{equation}
Here the first term $F^{}_{ab}$ in (\ref{F3}) is
the $U^{}_C(1)$ (Cartan)  field strength tensor, and (\ref{F+-}) involves
the $U^{}_C(1)$ covariant derivatives $\mathrm D_{A\, b}^\pm$ 
of the charged vector
fields $X^\pm_a$. These terms are clearly consistent with the 
$U^{}_C(1)$ multiplet structure.  The sole term that lacks
an obvious physical interpretation in terms of 
the $U^{}_C(1)$ multiplet, is the antisymmetric 
tensor $P^{}_{ab}$ in (\ref{F3}). We now proceed to interpret
it geometrically.

\vfill\eject

%
%
\section{Grassmannians and Spin-Charge Separation}
%
%

We shall now interpret the antisymmetric tensor $P^{}_{ab}$ in
(\ref{F3}). Explicitely we have
\begin{equation}
P_{ab} = \frac{i}{2} (X_a^+ X_b^- - 
X_b^+ X_a^-) = A^1_a A^2_b - 
A^1_b A^2_a
\la{grass}
\end{equation}
This is a real antisymmetric $4\times 4$ matrix that obeys
the quadratic relation
\begin{equation}
P_{12}P_{34} - P_{13}P_{24} + P_{23}P_{14} = 0
\la{quadr}
\end{equation}
In fact, using simple linear algebra one can show 
that {\it any} real $4\times 4$ antisymmetric 
matrix $P_{ab}$ that is subject to the condition (\ref{quadr}) 
can always be represented in the functional form (\ref{grass}) in terms 
of some two vectors $A^1_a$ and $A^2_b$. In projective 
geometry the relation 
(\ref{quadr}) is known as the {\it Klein quadric}. It 
describes the embedding of the real Grassmannian $G(4,2)$
in the five dimensional projective space $\mathbb R \mathbb  P^5$ 
as a degree four hypersurface. This Grassmannian
is the four dimensional 
manifold of two dimensional planes that are embedded 
in $\mathbb R^4$, and it can
be identified with the homogeneous space
\[
G(4,2) \simeq \frac{SO(4)}{SO(2) \times SO(2)}
\] 

It is convenient to describe the two dimensional planes 
in $\mathbb R^4$ in terms of a {\it zweibein}. For this
we introduce an orthonormal doublet $e_{a}^{\alpha}$ ($\alpha =1,2$)
\[
e_a^\alpha e_a^\beta = \delta^{\alpha \beta}
\]
that spans a generic two dimensional plane in $\mathbb R^4$. We can then 
represent the off-diagonal components of the gauge field as
\[
A_a^\alpha \ = \ {M^\alpha}_\beta e^\beta_a
\]
where ${M^\alpha}_\beta$ is a $2\times 2$ matrix. 
In terms of the complex combination
\[
e_a^{} = \frac{1}{\sqrt{2}} (e_a^1 + i e_a^2)
\]
we have
\begin{equation}
\begin{matrix}
e_a^{} e_a^{} = 0 \\
e^{}_a {\bar e}^{}_a = 1
\end{matrix}
\la{enorm}
\end{equation}
and we can write 
\begin{equation}
X_{a}^+
= A^1_a \pm i A^2_a =
\psi_1^{} e_a^{} + \psi_2^{} {\bar e}_a^{}
\la{decom}
\end{equation}
Here the $\psi^{}_{\alpha}$ are two arbitrary complex 
functions, they are linear combinations of the matrix elements 
of ${M^\alpha}_\beta$. 

When we substitute (\ref{decom}) into (\ref{grass}) we get
\begin{equation}
P^{}_{ab} = \frac{i}{2}(|\psi_1|^2 - |\psi_2|^2) \cdot 
(e^{}_a {\bar e}^{}_b
- e^{}_b {\bar e}^{}_a) \ = \ \frac{i}{2}\cdot \rho^2 \cdot 
t^{}_3 \cdot (e^{}_a {\bar e}^{}_b
- e^{}_b {\bar e}^{}_a)
\la{Ppsi}
\end{equation}
We have here introduced the three component unit vector
\begin{equation}
\vec t  \ = \ \frac{1}{\rho^2} 
\left( \begin{matrix} \psi^*_1 & \psi^*_2 \end{matrix} \right)
\vec \sigma \left( \begin{matrix} \psi^{}_1 \\ 
\psi^{}_2 \end{matrix} \right)
\ = \
\left( \begin{matrix} \cos \phi \cdot \sin \theta \\ \sin\phi \cdot 
\sin \theta \\ \cos \theta \end{matrix} \right)
\la{n}
\end{equation}
where we employ the following angular parametrization
\begin{equation}
\begin{matrix}
\psi^{}_1 & = & \rho \, e^{i\xi} \cos \frac{\theta}{2} \,
e^{-i\phi/2} \\
\psi^{}_2 & = & \rho \, e^{i\xi} \sin \frac{\theta}{2} \,
e^{i \phi/2}
\end{matrix}
\la{psipara}
\end{equation}

The representation (\ref{Ppsi}) 
has a motivation in terms of the properties of 
$P^{}_{ab}$. For this we consider the action of
the general linear group $GL(2,\mathbb R)$ on the coordinates
$A_{a}^{1}$ and $A_a^2$ that describe our generic two-plane
\[
\left( \begin{matrix} A_a^1 \\ A_a^2 \end{matrix} \right)
\ \buildrel {\mathrm G} \over
\longrightarrow \ 
\left( \begin{matrix} \alpha & \beta \\ \gamma & \delta 
\end{matrix} \right) 
\left( \begin{matrix} A_a^1 \\ A_a^2 \end{matrix} \right)
\]
Here
\[
\mathrm G \ = \  
\left( \begin{matrix} \alpha & \beta \\
\gamma & \delta \end{matrix} \right) 
\]
is the matrix realization of the $\mathrm G \in GL(2, \mathbb R)$ 
on the two-dimensional
plane in $\mathbb R^4$. This gives
\[
P_{ab} \  \buildrel \mathrm G \over
\longrightarrow \ (\alpha \delta - 
\beta \gamma) P_{ab}
\ = \ \det \mathrm G \cdot P_{ab}
\]
Thus $P^{}_{ab}$ supports a one-dimensional representation of 
$GL(2,\mathbb R)$, where the orbit is parametrized by 
the prefactor in (\ref{Ppsi})
\[
|\psi^{}_1|^2 - |\psi^{}_2|^2 = \rho^2 \cdot t^{}_3
\]
and the $\mathrm G $-action corresponds to a scaling of the
density $\rho$ according to
\begin{equation}
\rho \  \buildrel \mathrm G \over
\longrightarrow \ \sqrt{\det \mathrm G }\cdot \rho
\la{ggl2}
\end{equation}
Clearly, the volume and orientation
preserving subgroup $SL(2,\mathbb R)\in GL(2,\mathbb R)$
is an invariance
group of $P^{}_{ab}$. The maximal compact subgroup
$U^{}_I(1) \in SL(2,\mathbb R)$ is the internal 
invariance group of the decomposition 
(\ref{decom}). It acts on the fields according to
\begin{equation}
\begin{matrix}
e^{}_a & \to & e^{-i \lambda} e^{}_a \\
\psi^{}_1 & \to & e^{i\lambda}\psi^{}_1 \\
\psi^{}_2 & \to & e^{-i\lambda}\psi^{}_2
\end{matrix}
\la{intu1}
\end{equation}
The vector $\vec t$ in (\ref{n}) is invariant under 
the external  $U^{}_C(1)$ gauge transformation. 
The component $t_3$ is in addition invariant under the internal 
$U^{}_I(1)$ gauge transformation. But the remaining two
components transform nontrivially under the 
internal $U^{}_I(1)$. With 
\[
t_{\pm} = t_1 \pm i t_2
\]
the $U^{}_I(1)$ transformation sends 
\begin{equation}
t^{}_\pm = \frac{1}{2} ( t^{}_1 \pm i t^{}_2) \ \to \ 
e^{\mp 2i\lambda}
t^{}_\pm
\la{tpm}
\end{equation}
where the factor of $2$ reflects the fact that $\vec t$ is a bilinear
in $\psi_\alpha$. 

{\it A priori}, the {\it r.h.s.} of the decomposition 
(\ref{decom}) involves a total of
nine independent field degrees of freedom. These are the four 
real components of the functions $\psi^{}_\alpha$ and the five real 
components of the (complex) normalized vector $e^{}_a$. But due to 
the internal $U^{}_I(1)$ symmetry 
both sides of the decomposition (\ref{decom}) 
describe an equal number of eight independent field  degrees of freedom.
This  coincides with the number of independent components 
in the off-diagonal gauge
field and confirms that (\ref{decom}) yields a full, complete 
field decomposition of the off-diagonal components $X^{a}_\mu$ 
of the gauge field. 

According to (\ref{intu1}) the complex scalar fields
$\psi^{}_{1}$ and $\psi^{}_2$ are oppositely charged
with respect to the internal $U^{}_I(1)$ gauge group. The
vector field $e^{}_a$ is also charged {\it w.r.t.} 
the internal $U^{}_I(1)$ group while $A_a$ remains obviously 
intact under a $U^{}_I(1)$ transformation. The obvious choice
of a $U^{}_I(1)$ connection is the composite vector field
\begin{equation}
C_a = i \, {\bar e}^{}_b\partial_a e^{}_b
\la{defG}
\end{equation}
as it transforms according to
\[
C_a \to C_a + \partial_a \lambda
\]
under the internal $U^{}_I(1)$ transformation (\ref{intu1}).

We note that in terms of the explicit parametrization
(\ref{psipara}), the $U^{}_{I}(1)$ transformation 
(\ref{intu1}) sends
\[
\phi \to \phi - 2\lambda
\]

We also note that the connection (\ref{defG}) admits a
geometric interpretation as a spin connection that parallel 
transports the zweibein $(e^{1}_a, e^2_a)$.

According to (\ref{h}), the Cartan subgroup $U^{}_C(1)$ of the
$SU(2)$ gauge group acts on the complex coefficients
as follows,
\[
\psi^{}_{1,2} \ \buildrel {\mathit h} \over \longrightarrow \
e^{-i\omega} \psi^{}_{1,2}
\]
or in terms of the parametrization (\ref{psipara})
\[
\xi \to \xi - \omega
\]
while $e_a$ remains intact. Consequently we can interpret
the $SU(2)$ gauge field as the $U^{}_C(1)$ multiplet
\[
A_a^i \ \sim \ ( A_a, \psi^{}_{1} , \psi^{}_2,
e^{}_a )
\]
where $A_a$ is the $U^{}_C(1)$ gauge field, the complex scalar 
fields $\psi^{}_{\alpha}$ are
equally charged {\it w.r.t.} the $U^{}_C(1)$,
and the complex vector  field $e^{}_a$ is $U^{}_C(1)$ neutral. 

We can also interpret the gauge field as the $U^{}_I(1)$ multiplet
\[
A_a^i \ \sim \ (C_a, \psi^{}_{1}, \psi^{}_2, A_a)
\]
where $C_a$ is the $U^{}_I(1)$ gauge field, the complex scalar
fields are oppositely charged {\it w.r.t.} the $U^{}_I(1)$ and the vector
field $A_a$ is $U^{}_I(1)$ neutral.

Notice that the $\psi_{\alpha}$ are 
scalars and $e_a$ is a vector under $SO(4)$ rotations  ({\it a.k.a.} 
Lorentz transformations) in $\mathbb R^4$. Since $e_a$ is neutral
under the $U^{}_C(1) \in SU(2)$ gauge group while the $\psi_{\alpha}$
transform nontrivially, we conclude that the 
decomposition (\ref{decom}) 
entails a separation between the spin and the charge 
in the off-diagonal components $X_a^\pm$ of the $SU(2)$ gauge field. 
The spinless scalar fields $\psi_{\alpha}^{}$ describe 
the $U^{}_C(1)$ charge degrees of freedom of the $X_a^\pm$, 
and the $U^{}_C(1)$ neutral vector field $e_a$ 
describes their spin degree of freedom. 

The present
separation between the spin and the charge degrees of freedom 
in $X_a^\pm$ is quite analogous to the slave-boson decomposition 
of an (nonrelativistic) electron, widely employed 
in attempts to explain high-temperature superconductivity \cite{and2}.
There, the spin-charge separation is introduced as an alternative
to the BCS superconductivity. Instead of the Higgs effect 
in terms of the Cooper pairs, superconductivity
emerges when the analog of our variable $\rho$ forms a condensate,
\begin{equation}
<\! vac| \, \rho \, | vac \! > = \Delta \not= 0
\la{rhocond}
\end{equation}

\vfill\eject

%
%
\section{Electric and magnetic components}
%
%

Consider the tensor part of $P^{}_{ab}$ in (\ref{Ppsi}),
\begin{equation}      
{H}_{ab} = \frac{i}{2} (e^{}_a {\bar e}^{}_b - e^{}_b {\bar e}^{}_a )
\la{7gmn}
\end{equation}
We define its ``electric'' ($p_i$) and ``magnetic'' ($q_i$) components
in the usual manner, by setting
\begin{equation}
\begin{array}{cc}
p_i & =  H_{0i} = \frac{i}{2} (e^{}_0 e^*_i -
e^{}_i e^*_0) 
\\
q_i & = \frac{1}{2} \epsilon_{ijk}H_{jk} = \frac{i}{2}
\epsilon_{ijk} e^{}_j e^*_k 
\end{array}
\la{7eb}
\end{equation}
These two vectors are subject to the orthogonality relations
\begin{equation}
\begin{matrix}
\vec p \cdot \vec q = 0 \\
\vec p\cdot \vec p + \vec 
q \cdot \vec q = \frac{1}{4}
\end{matrix}
\la{ebort}
\end{equation}
Together with the Poynting vector
\[
\vec s = \vec p \times \vec q
\]
we then have an orthogonal triplet in $\mathbb R^3$. 

The
relation (\ref{7eb}) can also be inverted, the result is
\begin{equation}
\vec e =
\left( \begin{array}{c}
e_0 \\ e_1 \\ e_2 \\ e_3 \end{array} \right) =
e^{i\eta}  \left( \begin{array}{c}
\sqrt{2} |\vec p| \\ \frac{
2 \vec s + i \vec p}
{  \sqrt{2} |\vec p|}
\ \end{array} \right) \ \equiv \ e^{i\eta} \hat {\vec e}
\la{7epq}
\end{equation}
Here $\eta$ is an overall phase of $\vec e$. This phase is invisible 
to $\vec p$ and $\vec q$ since 
it does not contribute to the bilinear combination (\ref{7gmn}). 
But this phase is subject to the internal $U^{}_I(1)$ gauge transformation
(\ref{intu1}) which sends
\begin{equation}
\eta \to \eta - \lambda
\la{u1I}
\end{equation}
Thus the phase transformation (\ref{u1I}) determines a rotation between the
real and imaginary components $e^1_a$ and $e^2_a$ of the complex
vector $e^{}_a$ in $\mathbb R^4$. 

Now consider the action of $SO(4)$ rotations ({\it a.k.a.}
Lorentz transformations) on $e_a$. 
We are particularly interested in the effect of an infinitesimal 
$SO(4)$ (Lorentz) boost in a generic spatial direction 
$\varepsilon_i$. As a component of a four-vector, the
$e^{}_a$ should transform in the following manner
\begin{equation}
\begin{matrix}
\Lambda_{\varepsilon} \ e^{}_0 = - \varepsilon_i e^{}_i \\
\Lambda_{\varepsilon} \ e^{}_i = - \varepsilon_i  e^{}_0
\end{matrix}
\la{boost1}
\end{equation}
This is clearly a $SO(4)$ rotation,
it preserves the orthonormality relations (\ref{enorm}).

On the other hand, we expect that when we realize this 
boost transformation on the electric and magnetic components (\ref{7eb}) 
we should get the familiar results
\begin{equation}
\begin{matrix}
\delta_\varepsilon \ \vec p = 
\vec q \times \vec \varepsilon \\
\delta_\varepsilon \ \vec q 
= \vec p \times
\vec \varepsilon
\end{matrix}
\la{boost2}
\end{equation}
Curiously, we find that when we substitute this in the explicit
realization (\ref{7epq}) there is a {\it difference} 
between (\ref{boost1}) 
and (\ref{boost2}): If we compare the action of $\Lambda_{\varepsilon}
$ in (\ref{boost1}) with the
action of $\delta_\varepsilon$ on $e^{}_a$ which is defined
using (\ref{boost2}) and (\ref{7epq}), the two descriptions 
of the boost differ from each other by the (infinitesimal) phase
\[
(\Lambda_\varepsilon - \delta_\varepsilon) \ e^{}_a =
- i \Theta( \vec p, \vec q ; \vec \varepsilon ) \cdot
e^{}_a
\]
As a consequence the difference between the two 
Lorentz boosts is a (infinitesimal) 
shift in the angle $\eta$ in (\ref{7epq}) according to
\[
\eta \to \eta - \Theta( \vec p, \vec q ; \vec \varepsilon )
\]
This is an infinitesimal internal $U^{}_I(1)$ gauge rotation (\ref{u1I}).

Explicitely, we have
\begin{equation}
\Theta( \vec p, \vec q ; \vec \varepsilon ) \
= \
\frac{ \vec p \cdot \vec \varepsilon }{|\vec p|^2}
= \varepsilon^i \frac{\partial}{\partial p_i}
\ln {|\vec p |}
\la{cycle1}
\end{equation}
One can verify that this quantity obeys the one-cocycle condition 
\[
\delta_{\varepsilon_1} \Theta( \vec p, 
\vec q ; \vec \varepsilon_2 ) \ - \ 
\delta_{\varepsilon_2} \Theta( \vec p, \vec q ; 
\vec \varepsilon_1 )  \ = \ 0
\]
and as a consequence (\ref{cycle1}) is a {\it one-cocycle}. 

The result means
that the action of the boost $\delta_{\varepsilon}$ on the vector
field $e^{}_a$ determines a {\it projective} representation 
of the (Euclidean) Lorentz group. In particular, 
the vector field $\hat e_a$ 
in (\ref{7epq})
\[
\hat {\bf e} = e^{-i\eta} {\bf e}
\]
transforms under the projective representation
according to
\[
\delta_{\varepsilon} \hat {\bf e} = \Lambda_{\varepsilon} \hat {\bf e} +
i \Theta \, \hat {\bf e}
\]

We conclude that
the phase $\eta$ in (\ref{7epq}) 
is a non-trivial field degree of freedom. If we set $\eta = 0$
in (\ref{7epq}), a spatial boost will generate a nontrivial
$\eta$ determined by the one-cocycle (\ref{cycle1}). 
Moreover, since the cocycle depends only on the electric
component of (\ref{7gmn}) we propose that $\eta$ can be viewed
as a phase (angular) variable for electric circulation.

Consider the internal $U^{}_I(1)$ connection (\ref{defG}). 
It admits the following explicit realization
\begin{equation}
C_a \ = \ i \, {\bar  {\bf e } } \cdot \partial_a \vec e
\ = \ i \, \hat {\bar {\bf e}} \cdot \partial_{\mathit a} {\hat {\bf e}}
- \partial_{\mathit a }\eta \ = \ 2 | \vec q| \, ( \vec k \times \vec l \cdot
\partial_{\mathit a} \vec k) - \partial_{\mathit a} \eta \ = \ {\hat{
{\! \mathit C}}}_{\mathit a}  - \partial_{\mathit a} \eta
\la{Gamma}
\end{equation}
Here $\vec k$ and $\vec l$ are two mutually orthogonal 
unit vectors in the electric and magnetic 
directions respectively,
\begin{equation}
\begin{matrix}
\vec p & = & |\vec p| \vec k  & = &  
\frac{1}{2\sqrt{2}}\cos \vartheta \cdot \vec k  \\
\vec q & = & |\vec q| \vec l &  =  &
 \frac{1}{2\sqrt{2}} \sin \vartheta \cdot
\vec l \end{matrix}
\la{ebang}
\end{equation}
Since both $\vec k$ and $\vec l$ remain intact under both 
the external $U^{}_C(1)$ and the internal 
$U^{}_I(1)$ gauge transformations we conclude that the
vector field
\begin{equation}
{\hat{\! \mathit C}}_a \ = \ C_a + \partial_a \eta \ = \
i \, {\hat {\bar {\mathbf e}}} \cdot \partial_a \hat {\mathbf e}
\ = \ 2 | \vec q | \, ( \vec k \times \vec l \cdot
\partial_a \vec k) = \frac{2\, \vec p \cdot \partial_a \vec s}{\vec p^2}
\la{Kmu}
\end{equation}
is gauge invariant both under the external
$U^{}_C(1)$ and under the internal $U^{}_I(1)$
gauge transformations; The internal $U^{}_I(1)$ 
acts only on the phase variable $\eta$ in (\ref{Gamma}) 
according to (\ref{u1I}).

If we introduce the
normalization
\begin{equation}
\omega_a = - \frac{1}{2|\vec q|} \, \hat C_a = - 2 \vec k \times \vec
l \cdot \partial_a \vec k
\la{kiri1}
\end{equation}
we arrive at 
\begin{equation}
\partial_a \omega_b - \partial_b \omega_a = \vec l \cdot
\partial_a\vec l \times \partial_b \vec l
\la{kiri2}
\end{equation}
This is the pull-back of the volume two-form on $\mathbb S^2$
that also appears in the second term of (\ref{0fha}).
As a consequence (\ref{kiri1}) admits the geometric 
interpretation as the Kirillov one-form for the co-adjoint
orbit $\mathbb S^2 = SU(2)/U(1)$, in the magnetic 
direction of $\vec l$. 

The result (\ref{kiri2}) follows directly from the formal 
properties of the unit vectors $\vec k$ and $\vec l$. Alternatively,
it can be verified by using an explicit angular representation
of these vectors. If we denote by $\vec m$ the unit vector 
in the direction of the Poynting vector,
\[
\vec m = - \vec k \times \vec l
\]
we can represent the orthonormal triplet $(\vec k, \vec l, \vec m)$
as follows,
\begin{equation}
\begin{matrix}
\vec k & \hskip -0.3cm =  \cos \gamma \cdot
\vec u_x + \sin \gamma \cdot \vec u_y
\\
\vec l & \hskip -3.5cm  = \vec u_z \\ 
\vec m & = - \sin \gamma \cdot
\vec u_x + \cos \gamma \cdot \vec u_y
\end{matrix}
\la{lkm}
\end{equation}
where
\begin{equation}
\vec u_x \ = \ \left( \begin{matrix} \cos \alpha \cos \beta
\\ \sin \alpha \cos \beta \\ - \sin \beta \end{matrix} \right) 
\ \ \ \& \ \ \ \ \vec u_y \ = \
\left( \begin{matrix} - \sin \alpha \\ \cos \alpha \\ 0 
\end{matrix} \right) \ \ \& \ \ \ \ \ \vec u_z \ = \
\left( \begin{matrix} \cos \alpha \sin \beta \\ \sin \alpha 
\sin \beta \\ \cos \beta \end{matrix} \right)
\la{exyz}
\end{equation}
We then have
\begin{equation}
A_M = - \frac{1}{4} \omega = - \frac{1}{2} (
\cos \beta \cdot d \alpha + d \gamma)
\la{moncon}
\end{equation}
and
\begin{equation}
F_M = -\frac{1}{4}d \omega \ = \ - \frac{1}{2}
\sin \beta \cdot d\alpha \wedge d\beta
\la{monfie}
\end{equation}
In (\ref{moncon}) we recognize the connection one-form $A_M$
and in (\ref{monfie}) the curvature two-form $F_M$ (magnetic field)
of the Dirac magnetic monopole. 

For (\ref{Kmu}) we get from (\ref{ebang})
\begin{equation}
{\hat{\! \mathit C}} = - \sqrt{2} \sin \vartheta \cdot
[ \cos \beta \cdot d\alpha + d\gamma]
\la{maglim}
\end{equation}
This vanishes  when we go to 
the purely electric limit $\vartheta =0$, and 
coincides with (twice) the Dirac monopole connection (\ref{moncon})
when $\vartheta = \pi/4$ and the strength of the 
electric and magnetic fields are equal.

\vfill\eject

%
%
\section{A Higgs effect}
%
%

The three components of the unit vector $\vec t$ that we have 
introduced in (\ref{n}) are bilinear in the complex
functions $\psi^{}_1$ and $\psi^{}_2$. Since these functions
are $SO(4)$ {\it a.k.a.} Lorentz scalars, the unit 
vector $\vec t$ is also a Lorentz scalar. But its 
components $t^{}_{\pm}$ are not invariant under the 
internal $U^{}_I(1)$ gauge 
transformations. 

If instead we introduce a new three component
unit vector $\vec n$ such that
\[
\begin{matrix} n_\pm \\ n^{}_3 \end{matrix} \ \begin{matrix} = \\
= \ \end{matrix} \
\begin{matrix}
e^{2i\eta} t_\pm \\ t^{}_3 \end{matrix}
\]
then this new unit vector is 
invariant under both the external
$U^{}_C(1)$ and the internal $U^{}_I(1)$
gauge transformations. Explicitely, in terms of the
angular variables in (\ref{n}) we have
\begin{equation}
\vec n = \left( \begin{matrix} \cos (\phi + 2\eta) \cdot \sin \theta \\
\sin (\phi + 2\eta) \cdot \sin \theta
\\ \cos \theta \end{matrix} \right)
\la{defn}
\end{equation}
Here
\[
\phi + 2\eta
\]
is a $U^{}_{C}(1) \times U^{}_I(1)$ invariant 
combination of the $U^{}_I(1)$ dependent variables $\phi$ and $\eta$.

In the sequel we shall propose that $\hat{\mathbf e}_a $ in (\ref{7epq}),
$\hat C_a $ in (\ref{Kmu}) and $\vec n$ are the obvious
$U^{}_C(1) \times U^{}_I(1)$
gauge invariant variables for describing the Yang-Mills theory.
For this we note that the relation (\ref{Kmu}) between $C_a$ and 
$\hat C_a$ is a version of the familiar (linear) Higgs relation 
between the $U(1)$ gauge vector field and the gauge 
invariant (massive) vector field in the standard
abelian Higgs model:

In the case of conventional Higgs mechanism, a $U(1)$ gauge field combines 
with the gradient of the phase of a complex scalar field into a 
gauge invariant vector field. The modulus of the 
complex scalar field remains as an additional, independent gauge 
invariant field variable. 
When this modulus develops a nonvanishing
ground state expectation value the gauge invariant 
vector field becomes massive.

Here, the $U^{}_I(1)$ gauge field $C_a$ 
combines similarly with the phase $\eta$ of the complex vector 
field $\mathbf e_a$ into the gauge invariant vector field $\hat C_a$.
This leaves the vector field $\hat {\mathbf e}^{}_a$ in (\ref{7epq}) 
as an additional independent and $U^{}_I(1)$ invariant field variable. 
Furthermore, the transition
from $\vec t$ to the $U^{}_I(1)$ invariant $\vec n$ can be viewed 
as a {\it nonlinear} version of the Higgs mechanism. Note that
all of these three $U^{}_I(1)$ invariant 
field variables are also invariant under
the $U^{}_{C}(1)$ gauge transformations.

We remind that due to the presence of
the one-cocycle (\ref{cycle1}) the 
vector fields $\hat C_a$ and $\hat {\mathbf e}_a$
are not $SO(4)$ vectors. Nor are the $\pm$-components of 
$\vec n$ scalars under $SO(4)$. Instead, all of these
$U^{}_I(1)$ gauge invariant quantities transform under a projective 
representation of the spatial $SO(4)$ ({\it a.k.a.} Lorentz) group. 

The breaking of the Lorentz invariance by the one-cocycle has
important physical consequences. For this we note that the 
two components $A^1_a$ 
and $A^2_a$ of the $SU(2)$ gauge field appear symmetrically in the
Yang-Mills Lagrangian, they can be exchanged by a global gauge
transformation. Consequently we can expect that 
in terms of the spin-charge separated variables the Lagrangian 
should display a similar global symmetry between the two complex scalar 
fields $\psi^{}_1$ and $\psi^{}_2$. This symmetry
should translate into a global $O(3)$ rotation invariance
when represented in terms of the unit vector $\vec n$ 
in (\ref{defn}); See the first two terms in (\ref{0fha}).  But
the presence of such a global $O(3)$ symmetry poses a problem. When we
select a ground state direction of $\vec n$ we break the 
global $O(3)$ invariance explicitly, in a manner which in general
leads to two {\it a priori} massless Goldstone bosons. 

The violation of Lorentz invariance by the one-cocycle
appears to remove the Goldstone bosons: Since we have no reason to expect 
that the ground state of the theory violates Lorentz invariance
and since $n^{}_3$ is the sole Lorentz invariant component of
$\vec n$, the only possible Lorentz invariant ground state 
direction for $\vec n$ is
\[
\vec n \ \to \
\pm \left( \begin{matrix} 0 \\
0 \\ 1 \end{matrix} \right) \ \equiv \ \hat {\vec z}
\]
{\it Any } nonvanishing ground state value for the components
$n^{}_\pm$ breaks
Lorentz invariance due to the presence of the one-cocycle. 

In particular, we conclude that at large spatial distances
the unit vector $\vec n$ should become asymptotically parallel 
with the $z$-axis,
\begin{equation}
\vec n \ \buildrel{|x| \to \infty} \over {\longrightarrow} \  
\pm \hat {\vec z}
\la{defnasym}
\end{equation}
Alternatively, in terms of the vector $\vec t$ the only 
$U^{}_{I}(1)$ gauge invariant asymptotic ground state direction
is 
\[
\vec t \to \pm \hat {\vec z}
\]
since any other asymptotic direction violates the internal
$U^{}_I(1)$ gauge invariance.

\vfill\eject

%
%
\section{Yang-Mills in Gauge Invariant
Spin-Charge Variables}
%
%

We now proceed to inspect how the 
separation between the spin and the charge manifests
itself in the Yang-Mills Lagrangian. 
We start by recalling the tree-level gauge fixed 
Euclidean space Yang-Mills Lagrangian
\begin{equation}
L^{}_{YM} \ = \ \frac{1}{4} (F^i_{ab})^2 + \frac{\xi}{2}
|\mathrm D^+_{A \, a} X^{+}_{a}|^2 + L^{}_{ghost}
\la{YMclas}
\end{equation}
Note that we have here introduced a gauge fixing term only 
for the off-diagonal components $X^\pm_a$ of the gauge field.
For reasons that will eventually become transparent,
we do not introduce any gauge fixing term
in the direction of the abelian subgroup $U^{}_C(1)
\in SU(2)$. The last term $L^{}_{ghost}$ denotes the
ghost contribution. In the sequel its explicit form will not be 
of importance to us. We only need to observe that it is entirely
independent
of the gauge fixing parameter $\xi$ \cite{kondo}, \cite{anie}.

In our approach we do {\it not} introduce
decomposed variables in the path integral. That would only
lead to unnecessary complications. Instead we 
propose that the appropriate stage to implement 
the spin-charge separation is at the level 
of the effective Yang-Mills action which has been computed in 
the covariant background field formalism.  This effective action
accounts for all quantum fluctuations in the gauge field.  
But since its explicit form is not available beyond a few leading terms 
in a loop expansion, we need to resort to an indirect analysis.

By general arguments of gauge invariance we can expect that the 
full effective action is a functional of the
{\it background} field strength tensor $F^i_{ab}$ and its
background 
covariant derivatives. In the low momentum infrared limit we can 
ignore the derivative contributions, hence in this limit 
the effective action involves only the field strength tensor. 
Since the full result is unknown to us, for simplicity 
we proceed by considering the infrared limit only 
in its lowest order. This limit coincides with the classical 
Lagrangian (\ref{YMclas}), but excluding the ghost contribution.
Consequently our starting point will be the classical Yang-Mills
Lagrangian (\ref{YMclas}). Indeed,
the classical Lagrangian {\it should} be an important 
ingredient of the full quantum action. We now proceed
to subject it to the 
separation between the spin and the charge.

When we introduce (\ref{F3}) and (\ref{F+-}) we find for the
classical Yang-Mills Lagrangian
\[
L^{}_{YM} \ = \ \frac{1}{4} (F_{ab}^i)^2 \ + \  \frac{\xi}{2}
|\mathrm D^+_{A\, a} X^+_a|^2
\]
\begin{equation}
= \
\frac{1}{4}( F_{ab} + 2 \rho^2 n^{}_3 
H_{ab})^2 \ +  \ \frac{1}{2} |\mathrm D^+_{A\, a}X^+_b|^2 
+ \frac{3}{8} (1-n_3^2) \rho^4 - \frac{3}{8} \rho^4 
 + \  \frac{\xi - 1}{2}|\mathrm D^+_{A\, a} X^+_a|^2 
\la{YMact1}
\end{equation}
The reason why we present the third and fourth terms in (\ref{YMact1})
in this particular manner becomes evident as we proceed.

Note that we have here overlooked 
a surface contribution that originates from the
difference
\[
\mathrm D^+_{A\, a}X^+_b  \mathrm D^-_{A\, b} 
X^-_a \ - \  \mathrm D^+_{A\, a} X^+_a
\mathrm D^-_{A\, b}X^-_b
\]
Explicitely, the surface contribution 
is
\begin{equation}
\partial_a \left\{ \frac{1}{2} 
\left[ X^+_a \mathrm D^-_{A\, b} X^-_b \ + \
X^-_a \mathrm D^+_{A\, b} X^+_b \right] - \frac{1}{2} 
\partial_b (
X^+_a X^-_b ) \right\}
\la{surf}
\end{equation}

We first observe that in (\ref{YMact1}) 
there are two particularly interesting values for
the gauge fixing parameter $\xi$. These are the value
$\xi = 1$, and the limit $\xi \to \infty$. 

If  we select 
$\xi=1$ the last term in (\ref{YMact1}) becomes absent. 
In particular, when $\xi = 1$ there are no terms present in the 
Lagrangian barring the surface term,
where the Lorentz index in the off-diagonal components 
$X^{\pm}_a$ becomes contracted with the Lorentz indices that 
are carried by the other quantities, such as the derivative 
operator $\partial_a$. For  $\xi = 1$ the 
Lorentz indices in $X^{\pm}_a$ are {\it only} contracted internally 
between different contributions of the $X^{\pm}_a$.
 
Since the ghost Lagrangian is independent of $\xi$, 
by arguments of gauge invariance we expect that this property persists 
to all orders of perturbation theory. In particular  we 
expect that in the full $\xi = 1$ quantum effective 
action the Lorentz indices carried by the (background) 
fields $X^\pm_a$ are only contracted internally, between
different contributions of $X^\pm_a$.
This indicates that we can (crudely) analyze the feasibility 
of the spin-charge 
separation by considering the Lagrangian (\ref{YMact1}), 
with $\xi = 1$ and ignoring the ghost contributions. As a consequence 
we limit our interest to only the following four terms,
\begin{equation}
L^{}_{YM} = \frac{1}{4}( F_{ab} + 2 \rho^2 n^{}_3 
H_{ab})^2  + \  \frac{1}{2} |\mathrm D^+_{A \,a} X^+_b|^2 
+ \frac{3}{8} (1-n_3^2) \rho^4 - \frac{3}{8} \rho^4 
\la{YMact}
\end{equation}

Similar conclusions can be drawn in the gauge that emerges
when we send $\xi \to \infty$. In this limit
we obtain in addition the maximal abelian gauge (MAG) condition
\begin{equation}
(\partial_a \pm i A_a) X^\pm_a \ = \ \mathrm D^\pm_{A\, a} 
X^{\pm}_{a} \ = \ 0
\la{mag}
\end{equation}
It is known \cite{kondo},
\cite{zakh} that this gauge condition is also
the (Euler-Lagrange) variational equation that 
describes the gauge orbit extrema of the following quantity,
\begin{equation}
\mathcal R = \int d^4x \ X^+_a X^-_a
\la{R}
\end{equation}
In particular,
\[
\rho^2 = |\psi^{}_1|^2 + |\psi^{}_2|^2 = X^+_a X^-_a
\]
where $\rho$ is the density that we have introduced in (\ref{psipara}). 
Notice that with (\ref{mag}) the first 
two terms in the surface contribution (\ref{surf}) vanish.

The variable $\rho$ has an interpretation as a condensate;
see (\ref{rhocond}). In the context of the maximal abelian gauge 
this interpretation of $\rho$ has been discussed widely in the 
literature \cite{kondo}, \cite{zakh}: The extrema values of $\rho$ 
on the gauge orbit are obviously gauge invariant and 
according to (\ref{R}) correspond to gauge field configurations 
that are subject to the MAG gauge condition (\ref{mag}) and we refer
to \cite{kondo}, \cite{zakh} for further discussion.

From (\ref{ggl2}) we also conclude that selecting the extrema value 
of $\rho$ breaks the Grassmannian $GL(2,\mathbb R)$ into 
$SL(2,\mathbb R)$.

We now proceed to analyze the Lagrangian (\ref{YMact}). In the present
section our goal will be to represent it in terms of the 
$U^{}_C(1) \times U^{}_I(1)$ gauge
invariant variables. We shall find that this can be achieved by
a change of variables, with no need to any additional explicit gauge fixing.
This will justify
{\it a posteriori} why in (\ref{YMclas}) we have introduced the gauge fixing
term only for the off-diagonal $X^\pm_a$.

We first consider the second term in (\ref{YMact}). 
Using (\ref{n}) we can write 
this term as
\begin{equation}
|\mathrm D^+_{A \, a} X^+_b|^2 = 
|\mathrm D^C_{A\, a} \psi_1|^2 + |\mathrm D^C_{A\, a} \psi_2|^2 + \rho^2
|\mathrm D^C_{A\, a} e_b|^2 + \frac{1}{2} \rho^2 t_+ 
(\bar{\mathrm D}^C_{A\,_a} \bar e_b)^2
+ \frac{1}{2} \rho^2 t_-  (\mathrm D^C_{A\, a} e_b)^2
\la{YM2nd}
\end{equation}
Here $\mathrm D^C_{A\, a}$ is the following 
$U^{}_C(1) \times U^{}_I(1)$ 
covariant derivative,
\[
\begin{matrix}
\mathrm D^C_{A\, a} \psi^{}_1 & = &  (\partial_a  + i A_a  -
i C_a ) \psi^{}_1
\\
\mathrm D^C_{A\, a}\psi^{}_2 & = & (\partial_a + i A_a  +
i C_a) \psi^{}_2
\\
\mathrm D^C_{A\, a} e^{}_b & = & \hskip 0.4cm (\partial_a + i C_a) e_b
~~~~~~~~~~~~
\end{matrix}
\]
Note that even though the $t_\pm$ are not invariant 
under the internal $U^{}_I(1)$ gauge transformations 
(\ref{tpm}), since $\mathrm D^C_{A\, a} e_b$ transforms according to
\[
\mathrm D^C_{A\, a} e_b \ \to \ e^{-i\lambda} \mathrm D^C_{A\, a} e_b
\]
the Lagrangian (\ref{YM2nd}) is gauge invariant 
under both $U^{}_I(1)$ and $U^{}_C(1)$ gauge transformations.

We introduce the $U^{}_C(1) \times U^{}_I(1)$ invariant
supercurrent \cite{omat2}, \cite{babaev}
\[
\hskip 1.5cm
J^{}_a = \frac{i}{2\rho^2} \{ \psi^*_1  \mathrm D^C_{A \, a}
\psi_1 - \psi^{}_1 \bar {\mathrm D}^C_{A \, a} \psi^*_1 + 
 \psi^*_2  {\mathrm D}^C_{A\, a}
\psi_2 - \psi^{}_2 \bar {\mathrm D}^C_{A \, a} \psi^*_2
 \} 
\]
From this we can solve for $A_a$ in favor of 
$J_a$. The result is
\[
A_a = - J_a + \frac{i}{2\rho^2} \{ \psi^*_1 \overleftrightarrow{
\partial}_{\!\!\! a}
\, \psi^{}_1 + \psi^*_2 \overleftrightarrow{\partial}_{\!\!\! a} \, \psi_2 \}
+ n_3 \cdot C_a
\]
When we substitute this in (\ref{YM2nd}) 
we get for the first two terms 
\begin{equation}
|\mathrm D^C_{A\, a} \psi_1|^2 + |\mathrm D^C_{A\, a} 
\psi_2|^2 = (\partial_\mu \rho)^2
+ \frac{1}{4} \rho^2 (\mathrm D^{\hat C}_{\, a} \vec n )^2 + \rho^2 J_a^2
\la{1st2}
\end{equation}
Here we have defined the covariant derivative operator
\begin{equation}
(\mathrm D^{\hat C}_{\, a})^{ij} = \delta^{ij} \partial_a + 
2 \epsilon^{ij3} \hat C_a
\ \ \ \ \ \ \ (i,j=1,2,3)
\la{nabladef}
\end{equation}
Note in particular that the middle term in the {\it r.h.s.} of 
(\ref{1st2}) is Lorentz invariant even though both the components
$n_\pm$ and the
connection $\hat C_a$ 
transform according to a projective representation of $SO(4)$.
The covariant derivative (\ref{nabladef}) 
compensates for the lack of $SO(4)$ invariance ({\it a.k.a.} 
Lorentz invariance) in the $i=1,2$ components of $\vec n$.

With (\ref{1st2}) we have achieved our goal, in the sense that 
the {\it r.h.s.} of (\ref{1st2}) involves only quantities 
which are $U^{}_C(1)\times U^{}_I(1)$ invariant.

We now proceed to the third term in (\ref{YM2nd}). For
this we get
\[
\rho^2 |(\partial_a + i C_a) \vec e|^2 \ = \
\rho^2 |(\partial_a + i \hat C_a) \hat {\vec e}|^2 \ = \
\frac{\rho^2}{2} \{( \partial_a \vec p)^2 +
(\partial_a \vec q)^2\}
\]
\begin{equation}
= \ \frac{\rho^2}{16} \left\{  \cos^2 \vartheta \cdot
(\partial_a \vec k)^2 + \sin^2 \vartheta \cdot (\partial_a \vec l)^2 
+ (\partial_a \vartheta)^2 
\right\}
\la{2ndeb}
\end{equation}
Clearly, this involves only manifestly $U^{}_C(1) \times U^{}_I(1)$
invariant quantities. 

We observe that there is the following 
apparent structural similarity between a contribution
to the second and third 
terms in the {\it r.h.s.} of
(\ref{1st2}), and the {\it r.h.s.} of (\ref{2ndeb}),
\[
\begin{matrix}
(\partial_a \vec n )^2 & \leftrightarrow & 
\cos^2 \vartheta \cdot
(\partial_a \vec k)^2 + \sin^2 \vartheta \cdot (\partial_a \vec l)^2 
\\
J_a^2 & \leftrightarrow &  (\partial_a \vartheta)^2
\end{matrix}
\]
In \cite{omat2} it has been suggested that this structural
similarity can be interpreted in terms of an electric-magnetic 
duality. Here we propose that it suggests a duality between
the spin and the charge.

The last two terms in (\ref{YM2nd}) can also be represented 
in terms of $U^{}_C(1) \times U^{}_I(1)$ gauge 
invariant variables as follows,
\[
\frac{1}{2} \rho^2 t^{}_+ ({ {\mathrm D}}^C_{A\, a}  {\bar e}^{}_b)^2
= \frac{1}{2} \rho^2 n^{}_+ (\partial_a {\hat {\bar e}}_b)^2
\]
\begin{equation}
= \frac{1}{128}  \frac{ \rho^2 n^{}_+}{|\vec s |^2} 
<\partial_a (\vec p +
\vec q) , \vec p - \vec q -
4\sqrt{2} \, i \vec s> \cdot < \partial_a (\vec p -
\vec q) , \vec p + \vec q -
4\sqrt{2} \, i \vec s>
\la{3dr}
\end{equation}
\[
\frac{1}{2} \rho^2 t^{}_- (\mathrm D^C_{A\, a}e^{}_b)^2
= \frac{1}{2} \rho^2 n^{}_- (\partial_a {\hat e}_b)^2 =
\]
\begin{equation}
= \frac{1}{128}  \frac{ \rho^2 n^{}_-}{|\vec s|^2} 
< \partial_a (\vec p +
\vec q) , \vec p - \vec q +
4\sqrt{2} \, i \vec s> \cdot < \partial_a (\vec p -
\vec q) , \vec p + \vec q +
4\sqrt{2} \, i \vec s>
\la{4th}
\end{equation}
Shortly we shall argue that these two terms admit a geometrical 
interpretation in the Grassmannian framework. However,
prior to this we consider the remaining contributions to the 
Yang-Mills Lagrangian.

We proceed with the first term in (\ref{YM2nd}). 
When we eliminate $A_a$ in favor of the supercurrent $J_a$ 
we get for this term
\begin{equation}
\frac{1}{4} (F_{ab} + 2\rho^2 t_3 H_{ab} )^2
\ = \frac{1}{4}( L_{ab} + M_{ab} - n^{}_3 K_{ab} -
2\rho^2 n_3 H_{ab} )^2
\la{GHP}
\end{equation}
Here
\begin{equation}
\begin{matrix}
L_{ab} & = & \hskip -0.5cm \partial_a J_b - \partial_b J_a \\
M_{ab} & = & \hskip 0.25cm 
\frac{1}{2} \, \vec n \cdot \mathrm D^{\hat C}_{ \, a} \vec n \times
\mathrm D^{\hat C}_{ \, b} \vec n  \\
K_{ab} & = & \hskip -0.5cm 
\partial_a \hat C_b - \partial_b \hat C_a
\end{matrix}
\la{GH}
\end{equation}
In particular, (\ref{GHP}) and (\ref{GH}) involve only quantities
which are explicitely $U^{}_C(1)$ and
$U^{}_I(1)$ invariant. The covariant derivative (\ref{nabladef}) 
ensures that all quantities are also 
independently $SO(4)$ (Lorentz) invariant.

We note that we can write the second and third
terms in the {\it r.h.s.} of (\ref{GHP}) as follows,
\begin{equation}
n^{}_3 (\partial_a \hat C_b - \partial_b \hat C_a)
- \frac{1}{2} \vec n \cdot \mathrm D^{\hat C}_{ \, a} \vec n \times
\mathrm D^{\hat C}_{ \, b} \vec n \ = \ \partial_a [ n_3 \hat C_b ]
- \partial_b [ n_3 \hat C_a  ] - \frac{1}{2}
\vec n \cdot \partial_a \vec n \times \partial_b \vec n
\la{thtensor}
\end{equation}
The structure in (\ref{thtensor}) is
reminiscent of the 't~Hooft tensor \cite{thooft}. The last
term is the pull-back of the volume two-form on $\mathbb S^2$, and 
if we introduce the corresponding Kirillov one-form (\ref{kiri1})
\[
- \frac{1}{2}
\vec n \cdot \partial_a \vec n \times \partial_b \vec n
\ = \ \partial_a Q_b - \partial_b Q_a
\]
we can combine the first three terms in the {\it r.h.s.} of
(\ref{GHP}) into
\[ 
L_{ab} + M_{ab} - n^{}_3K_{ab} = \partial_a
(J_b - n_3 \hat C_b  - Q_b) - \partial_b (J_a -n_3 \hat C_a - Q_a)   
\]

\vskip 0.5cm
In summary, when we combine our results we find that in terms of the 
spin-charge separated variables the Yang-Mills Lagrangian
has the following $U^{}_C(1) \times U^{}_I(1)$ 
invariant form
\[
L^{}_{YM} =  \frac{1}{4}{\mathcal F}_{ab}^2
+ \frac{1}{2}(\partial_a \rho)^2
+ \frac{1}{2}\rho^2  J_a^2 + \frac{1}{8}\rho^2  (\mathrm 
D^{\hat C}_{ \, a} 
\vec n)^2
+ \frac{\rho^2}{4} \left\{  ( \partial_a \vec p)^2 +
(\partial_a \vec q)^2 \right\}
\]
\begin{equation}
+ \frac{1}{4} \rho^2
\left\{ n^{}_+ ( \partial_a\hat{\bar e}_b)^2 + n^{}_- 
(\partial_a \hat e_b)^2 \right\}
+ \frac{3}{8} (1-n_3^2) \rho^4 - \frac{3}{8} \rho^4 
\la{YMfi}
\end{equation}
where
\[
\mathcal F_{ab} =  \partial_a J_b - \partial_b J_a
+ \frac{1}{2} \, \vec n \cdot \partial_{a} \vec n \times
\partial_{ b} \vec n -  \{ \partial_a (n_3\hat C_b) - 
\partial_b (n_3 \hat C_a) \} -
2\rho^2 n_3 H_{ab} 
\]
We find it noteworthy that the final Lagrangian (\ref{YMfi})
contains only
$U^{}_C(1)$ and $U^{}_I(1)$ invariant quantities, despite the
fact that in (\ref{YMact}) we have only 
introduced gauge fixing for the off-diagonal components $X^\pm_a$.
In particular, the $U^{}_C(1) \in SU(2)$ gauge invariance has been 
eliminated explicitely
by the introduction of gauge invariant variables.
This elimination of the $U^{}_C(1) \times U^{}_I(1)$ gauge
invariance has been at the expense of introducing variables  
$\hat C_a$, $\hat {\vec e}$ and $n_\pm$ 
which transform according to a projective representation of the
$SO(4)$ (Lorentz) group.  However, in (\ref{YMfi}) these variables
appear only in $SO(4)$ invariant combinations.

The final Lagrangian (\ref{YMfi}) has a very interesting structure.
It describes the interacting dynamics between a version of the
$O(3)$ nonlinear $\sigma$-model that one of us introduced in
\cite{fadpap} and the 
$G(4,2)$ Grassmannian nonlinear $\sigma$-model.

Clearly, the natural interpretation of the real scalar 
field $\rho$ is in terms of
a condensate. Since $\rho$ is a positive definite quantity
we can expect that it develops the non-vanishing 
ground state expectation value (\ref{rhocond})
that characterizes a material background in (\ref{YMfi}); see
\cite{kondo}, \cite{zakh}. 

Due to the presence of the third term in (\ref{YMfi}),
a nonvanishing $\Delta$ in (\ref{rhocond}) leads to an effective mass to the 
vector field $J_a$. As a consequence this vector field
is  subject 
to the Meissner effect. If we assume that at large 
distances we can ignore the contribution from
$J_a$, the remaining Lagrangian involves only variables
that describe a the present version of the $O(3)$ $\sigma$-model
and the $G(4,2)$ Grassmannian non-linear $\sigma$-model.

In the London limit where we replace $\rho$ by its ground state
expectation value (\ref{rhocond}), 
the version of the $O(3)$ nonlinear $\sigma$-model
that has been embedded in (\ref{YMfi}) has the following Lagrangian,
\begin{equation}
\frac{\Delta^2}{8}  (\mathrm D^{\hat C}_{ \, a} 
\vec n)^2 + \frac{1}{16} \left\{
\, \vec n \cdot \partial_{ a} \vec n \times
\partial_{ b} \vec n - 2 \cdot \{ \, \partial_a (n_3 \hat C_b) - 
\partial_b (n_3\hat C_a )\, \} \, \right\}^2 + \frac{3}{8}\Delta^2 (1-n_3^2) 
\la{fmfi}
\end{equation}
This is in close resemblance with the effective Lagrangian
(\ref{0fha}), which we have proposed previously could be
an effective model for $SU(2)$ Yang-Mills 
theory \cite{omat1}-\cite{omat3}.  
The difference stems from the fact that here the
$\pm$ components of the
order parameter $\vec n$ lack Lorentz invariance due to the
one-cocycle (\ref{cycle1}). The Lorentz invariance of the Lagrangian
is restored by the presence of the 
similarly Lorentz invariance violating
vector field $\hat C_a$. 

We note that the last term in (\ref{fmfi}) is an additional $O(3)$ 
symmetry breaking potential
term. It is Lorentz invariant since $n_3$ is the sole component
of $\vec n$ that is a scalar under Lorentz transformations.

The reason for the particular combination of the potential terms
that we have introduced in (\ref{YMact1}) (the third and fourth
terms) becomes now obvious: This combination ensures that the 
angular variable $\theta$ of $\vec n$ in the 
parametrization (\ref{defn}) acquires a positive mass term. 
This choice still leaves the potential term involving only $\rho$ 
with a negative sign. Eventually, this sign will 
also find an explanation.

The original model (\ref{0fha}) supports knotted closed strings
as stable solitons. The version (\ref{fmfi}) involves also the
dynamical gauge field $\hat C_a$ that restores Lorentz invariance
in the present case. It would be very interesting to understand
how the addition of this field affects the soliton structure 
of the theory.

We now proceed to identify the $G(4,2)$ nonlinear $\sigma$-model that 
has been embedded in (\ref{YMfi}). This embedding is determined
by the kinetic term (\ref{2ndeb}) that can be written as
\[
|(\partial_a + i C_a) \vec e|^2 = \frac{\rho^2}{2}\left\{ (\partial_a 
\vec p )^2 + (\partial_a \vec q)^2 \right\}
\]
This reveals the topological
\[
G(4,2) \ \sim \ \frac{SO(4)}{SO(2) \times SO(2)} \ \sim
\ \mathbb S^2 \times \mathbb S^2
\]
structure of the Grassmannian. We conclude that when we subject 
$\vec p$ and $\vec q$ to the two conditions (\ref{ebort}),
these two three-component vector fields
describe the four dimensional Grassmannian manifold $G(4,2)$.
Indeed, {\it a priori} the two vector fields $\vec p$ and $\vec q$ 
have six independent components. But due to the two 
conditions (\ref{ebort}) only four of the components are independent, 
and correspond to coordinates on 
the four dimensional Grassmannian manifold $G(4,2)$.

Now, we return to the two terms (\ref{3dr}) and (\ref{4th}) which
together with the
last term in (\ref{GHP}) describe the coupling
between the Grassmannian model and the $O(3)$ model. 
We argue that the Grassmannian contribution in the interaction
terms
(\ref{3dr}) and (\ref{4th}) can
be identified as the (anti)holomorphic one-form on the complex 
manifold $G(4,2)\sim \mathbb S^2 \times \mathbb S^2$. For this 
we introduce the explicit parametrization (\ref{lkm}), (\ref{exyz}).
When we specify to the magnetic limit where $\vartheta \to \pi/2$ 
in (\ref{ebang}) we find for the Grassmannian contribution in 
(\ref{3dr})
\begin{equation}
\frac{1}{|\vec s |}\cdot 
< \partial_a (\vec p +
\vec q) , \vec p - \vec q -
4\sqrt{2} \, i \vec s> \ 
\buildrel {\vartheta \to 0 } \over \longrightarrow \
- 2 e^{i\psi} ( \vec l + i \vec m) \cdot \partial_a \vec k \ = \
- 2e^{i\psi} ( d\beta + i \sin \beta d\alpha)
\la{hols2}
\end{equation}
This is the (unique) holomorphic one-form on the magnetic
two-sphere described by $\vec k$.
Similarly we find in the electric limit $\vartheta \to 0$
that the Grassmannian contribution to (\ref{3dr}), (\ref{4th})
yields the (anti)holomorphic one-form on the electric two-sphere
in $G(4,2) \sim \mathbb S^2 \times \mathbb S^2$ which is 
described by the unit vector 
$\vec l$. These observations endorse our proposal that the 
Grassmannian contributions in (\ref{3dr}), (\ref{4th}) 
engage the holomorphic and the anti-holomorphic one-forms 
on the complex manifold $G(4,2) \sim \mathbb S^2 \times \mathbb S^2$.

Notice that in the interaction terms (\ref{3dr}), (\ref{4th}) 
the phase $\psi$ in (\ref{hols2}) can be combined with the
phase of $n_\pm$ into
\[
\phi + 2\eta \ \to \ \phi + 2\eta + \psi
\]

Finally, for the last term in (\ref{GHP}) we have in the 
Lorentz invariant ground state where $n_3=\pm 1$
\[
\frac{1}{4} ( 2 \rho^2 n_3 H_{ab})^2 = \frac{1}{2} n_3^2 \rho^4 
\ \approx \ \frac{1}{2} \rho^4 
\]
When we compare this with the last term in (\ref{YMfi}) we conclude
that despite the negative sign of this term we have an
overall stability of the theory.

\vfill\eject

%
%
\section{Gauge Covariance}
%
%

Our description of the spin-charge separation employs the Pauli 
frame (\ref{pauli}), (\ref{pauli2}) that identifies the diagonal
matrix $\sigma^3$ with the direction
of the $U^{}_C(1)$ subalgebra in the $SU(2)$ Lie algebra. We now
proceed to show that the spin-charge separation is frame independent,
instead of $\sigma^3$ we can select the direction of the
Cartan subalgebra $U^{}_C(1)$ in $SU(2)$ in an arbitrary and space-time 
dependent manner. For this we introduce an {\it a priori} 
arbitrary ${\mathit g}(x) \in SU(2)$ and perform the conjugation
\[
\sigma^3 \ \buildrel {\mathit g} \over \longrightarrow \  {\mathit g} 
\, \sigma^3 {\mathit g}^{-1}
\ \buildrel {\tt def} \over = \ m_i\sigma^i \ = \ \vec {\hat m}
\]
\[
\sigma^{\pm} \ \buildrel {\mathit g} \over \longrightarrow \
{\mathit g} \, \sigma^{\pm} {\mathit g}^{-1} \ \buildrel {\tt def} \over = \
e_i^{\pm} \sigma^i \ \buildrel {\tt def} \over = \ \frac{1}{2}
(e_i^1 \pm i e_i^2)\sigma^i \ = \  \vec {\hat e}^\pm 
\]
Clearly, these matrices also satisfy the same 
algebra as (\ref{pauli}), with
$\vec {\hat m}$ the Cartan generator
\[
[ \vec {\hat m} , \vec {\hat e}^\pm ] = \pm 2 \vec {\hat e}^\pm 
\]
\[
[ \vec {\hat e}^+ , \vec {\hat e}^- ] = \vec {\hat m}
\]
The gauge transformation (\ref{2fin}) by the 
matrix $\mathit g$ maps (\ref{pauli2}) onto
\[
A_a \sigma^3 + X^+_a \sigma^+ +
X^-_a \sigma^- \ \to \ A_a \vec {\hat m} + X^+_a 
\vec {\hat e}^+ + X^-_a
\vec {\hat e}^- + {\mathfrak a}_a \vec {\hat m} + 
\frac{1}{2i} [ \partial_a \vec {\hat m} ,
\vec {\hat m}]
\]
where 
\[
{\mathfrak a}_a = - i tr [ \sigma^3 {\mathit g}^{-1} \partial_a {\mathit 
g}]
\]
We interpret this connection in the following manner: We define 
\begin{equation}
{\mathcal A}_a = {\mathcal A}^i_a \sigma^i \ = \
( A_a + {\mathfrak a}_a) \vec {\hat m} + \frac{1}{2i}
[ \partial_a \vec {\hat m} , \vec {\hat m}] \ = \ {\mathit g} 
A_a \sigma^3 {\mathit g}^{-1} + 2i {\mathit g} 
\partial_a {\mathit g}^{-1}
\la{cA}
\end{equation}
\[
= \ C_a  \vec {\hat m} + \frac{1}{2i}
[ \partial_a \vec {\hat m} , \vec {\hat m}] 
\]
and 
\begin{equation}
{\mathcal X}_a = {\mathcal X}^i_a \sigma^i \ = \
X^+_a \vec {\hat e}^+ + X^-_a
\vec {\hat e}^- = \mathit g (X^+_a \sigma^+ + X^-_a \sigma^-) 
{\mathit g}^{-1} 
\la{cB}
\end{equation}
Here ${\mathcal A}_a$ is the connection originally introduced by
Duan and Ge \cite{ge}, and subsequently by Cho \cite{chos}; see also
\cite{omat1}. 

We introduce a generic ${\mathit h}(x) \in SU(2)$ and redefine
\[ 
{\mathit g} \to {\mathit g \mathit h}
\]
This determines a transformation under which ${\mathcal A}$ transforms 
as a connection
\[
{\mathcal A} \ \to \ {\mathit h}{\mathcal A} {\mathit 
h}^{-1} + 2i {\mathit h} d {\mathit h}^{-1}
\]
while ${\mathcal X}$ transforms as a tensor,
\[
{\mathcal X} \ \to \ {\mathit h} {\mathcal X} {\mathit h}^{-1}
\]

In (\ref{cA}), (\ref{cB}) we have superficially fourteen field degrees
of freedom. These are the four components of $C_a$, the 
eight components of $X^\pm_a$
and the two independent components of $\vec {\hat m}$. However, if
we impose the $h$-covariant condition \cite{shaba}
\begin{equation}
{\mathcal D}[{\mathcal A}]_a^{ij} {\mathcal X}^j_a = 0
\la{shab}
\end{equation}
this condition eliminates two of the field variables and
we are left with only the twelve independent components of a four
dimensional $SU(2)$ gauge field.

The condition (\ref{shab}) is a gauge covariant version of the 
maximal abelian gauge condition (\ref{mag}). Explicitely,
when we substitute (\ref{cA}) and (\ref{cB})  in
(\ref{shab}) and use the identity  
\[
m^i 
(\delta^{ij}\partial_a + \epsilon^{ikj}  
{\mathcal A}^k_a)
{\mathcal X}^j_a \ = \ 0
\]
we conclude that (\ref{shab}) is {\it equivalent} to the 
condition (\ref{mag}) for the original components $(A_\mu, 
X^\pm_\mu)$. In particular, when we choose $\vec {\hat m} \equiv 
\sigma^3$ we find that (\ref{shab}) reduces to (\ref{mag}) and we retain
all our previous results. This confirms that 
our separation between the spin and the charge in the
Yang-Mills Lagrangian is gauge covariant, independent of the direction
of $U^{}_C(1)$ in the $SU(2)$ gauge group. Furthermore, the
connection by Duan and Ge and by Cho acquires a role in the 
gauge covariantization of our formalism.

\vfill\eject

%
%
\section{Conformal Geometry}
%
%

The Yang-Mills Lagrangian has a number of attractive features
that become transparent when we present it in 
terms of the independent spin and charge variables. 
For example, the Lagrangian can be related to
a two-gap superconductor model \cite{anie}, and
$\rho$ admits also an independent interpretation as a 
gauge invariant condensate \cite{kondo}, \cite{zakh}. 

Here we shall 
propose an alternative interpretation of the Yang-Mills 
Lagrangian. We shall propose that $\rho$ can 
be viewed as the conformal scale of a conformally flat
metric tensor, and (\ref{YMfi}) describes the coupling between
matter fields and the Einstein-Hilbert gravity in the presence
of a nontrivial cosmological constant.

The version of the spin-charge separated Yang-Mills Lagrangian that we
shall employ is the following,
\[
L^{}_{YM} = L^{(1)}_{YM} +   L^{(2)}_{YM} + L^{(3)}_{YM} 
+ L^{(4)}_{YM} 
\]
where
\begin{equation}
L^{(1)}_{YM} \ = \ \frac{1}{4}\left \{ \partial_a J_b - \partial_b J_a
+ \frac{1}{2} \, \vec n \cdot \mathrm D^{\hat C}_{ \, a} 
\vec n \times \mathrm D^{\hat C}_{ \, b} \vec n - n^{}_3 
(\partial_a \hat C_b - 
\partial_b \hat C_a ) -
2 \rho^2 n_3 H_{ab}\right \}^2 
\la{ein1}
\end{equation}
\begin{equation}
\hskip -9.8cm L^{(2)}_{YM} \ = \
\frac{1}{2} \rho^2  J_a^2 + \frac{1}{8}
\rho^2   (\mathrm D^{\hat C}_{ \, a} \vec n)^2
\la{ein2}
\end{equation}
\begin{equation}
\hskip -5.5cm 
L^{(3)}_{YM} \ = \ \rho^2  |\mathrm D^C_{A \, a} e^{}_b|^2 + \frac{1}{2} 
\rho^2  t_+ ({\bar{\mathrm D}}^C_{A \, a} \bar e^{}_b)^2
+ \frac{1}{2} \rho^2  t_-  (\mathrm D^C_{A\, a} e^{}_b)^2
\la{ein3}
\end{equation}
\begin{equation}
\hskip -7.5cm L^{(4)}_{YM} \ = \ \frac{1}{2} (\partial_a \rho)^2 + 
\frac{3}{8} (1-n_3^2) \rho^4  
-   \frac{3}{8}  \rho^4  
\la{ein4}
\end{equation}
Our goal is to write these terms in a manifestly covariant manner,
with the conformally flat metric tensor
\begin{equation}
g^{}_{\mu\nu} \ = \ \left( \displaystyle{ \frac{\rho}{\Delta}} \right)^2
\delta^{}_{\mu\nu}
\la{met1} 
\end{equation}
Here $\Delta$ is a constant with dimensions of mass:
Since $\rho$ has dimensions of mass we need to introduce 
$\Delta$ so that the components of the metric tensor acquire their 
correct dimensionality. The obvious choice is to identify 
$\Delta$ with the vacuum expectation
value of the condensate $\rho$ according to (\ref{rhocond}).

We introduce the vierbein
\begin{equation}
g^{} _{\mu\nu}
\ = \ \delta_{ab}\,
{E^a}_\mu
{E^b}_\nu 
\la{met1a}
\end{equation}
where
\[
\delta^{ab} \ = \ g^{\mu\nu} {E^a}_\mu {E^b}_\nu
\]
and the vierbein ${E^a}_\mu$ is given explicitely  by
\[
{E^a}_\mu = \frac{\rho}{\Delta}
{\delta^a}_\mu 
\]
with 
\[
{E^a}_\mu {E_b}^\mu = 
\delta^a_{\, \, \, \, b}
\]

The Christoffel symbol of the metric (\ref{met1}) is
\begin{equation}
\Gamma^\mu_{\nu \sigma} = \frac{1}{2} g^{\mu\eta}(
\partial_\nu g^{}_{\eta \sigma} + \partial_\sigma g{}_{\eta \nu } -
\partial_\eta g^{}_{\nu\sigma  } ) = 
\frac{1}{4} \{ \, \delta^\mu_\sigma \delta^\tau_\nu
+ \delta^\mu_\nu \delta^\tau_\sigma - \delta^{\mu\tau} 
\delta_{\nu\sigma} \, \} \partial_\tau \ln \sqrt{g}
\la{chris}
\end{equation}
where
\[
\sqrt{g} =  \left( \displaystyle{ \frac{\rho}{\Delta}} \right)^4
\]

The spin connection is defined by demanding 
covariant constancy of the vierbein,
\[
\partial_\mu {E_a}^\nu + \Gamma_{\mu\lambda}^\nu {E_a}^\lambda -
\omega^{\, \, b}_{\mu \, \, a} {E_b}^\nu = 0
\]
This gives
\begin{equation}
\begin{matrix}
\omega^{\, \, a}_{\mu \, \, b} = & {E^a}_\nu \nabla_\mu {E_b}^\nu \ = \
{E^a}_\nu( \partial_\mu {E_b}^\nu + \Gamma^\nu_{\mu\lambda} {E_b}^\lambda)
\\
\hskip 0.85cm = & - {E_b}^\nu \nabla_\mu {E^a}_\nu \ = \
- {E_b}^\nu ( \partial_\mu {E^a}_\nu - \Gamma^\lambda_{\mu\nu} {E^a}_\lambda)
\end{matrix}
\la{spinc}
\end{equation}
In these relations we also indicate how the covariant 
derivative $\nabla_\mu$ acts on the
vector and co-vector fields.

Explicitely we get from the metric tensor (\ref{met1}), (\ref{met1a}) 
for the spin connection
\[
\omega^{\, \, a}_{\mu \, \, b} \ = \ \frac{1}{4} \{ \, 
{\delta^a}_\mu {\delta_b}^\sigma -
\delta_{bd} \, {\delta^d}_\mu \, \delta^{ac} {\delta_c}^\sigma \,
\} \partial_\sigma
\ln \sqrt{g}
\]

We employ the vierbein ${E^a}_\mu$ and the complex Grassmannian
zweibein (\ref{enorm}) to introduce the following 
complex zweibein
\[
\begin{matrix}
{\mathbbmss e}^{}_\mu = {E^a}_\mu  e^{}_a \\
{\bar\ce}_\mu = {E^a}_\mu e^*_a
\end{matrix}
\]
This zweibein is then normalized {\it w.r.t.} the metric $g_{\mu\nu}$
according to
\[
\begin{matrix}
g^{\mu\nu} {\mathbbmss e}^{}_\mu \bar{\mathbbmss e}^*_\nu \ = \ 1 \\
g^{\mu\nu} {\mathbbmss e}^{}_\mu {\mathbbmss e}^{}_\nu 
= g^{\mu\nu} \bar{\mathbbmss e}^*_\mu \bar{\mathbbmss e}^*_\nu = 0
\end{matrix}
\]
When we push forward the spin connection into
\[
{\omega}^{\,\, a}_{\mu \, \, b} \ \to \
{\omega}^{\, \, \lambda}_{\mu \,\, \nu} \ 
= \  {E_a}^\lambda \, 
\omega^{\, \, a}_{\mu\,\, b} {E^b}_\nu
\]
we can  introduce the generally covariant version
of the connection (\ref{defG}), 
\begin{equation}
{\mathcal C}_\mu  \ = \
i \bar\ce^\sigma ( \partial_\mu \ce^{}_\sigma - \Gamma^\lambda_{\mu\sigma}
\ce^{}_\lambda + \omega^{\, \, \lambda}_{\mu \,\, \sigma}\ce^{}_\lambda )
\ = \ i {\bar \ce}^\sigma \nabla_\mu \ce_\sigma
+ i \bar\ce^{\lambda} \omega^{\,\, \sigma}_{\mu \, \, 
\lambda}  \ce^{}_\sigma
\la{covC}
\end{equation}
and when we twist the covariant derivative operator with (\ref{covC}),
\begin{equation}
\nabla^{\mathcal C}_\mu \ = \ \nabla_\mu + i \mathcal C_\mu
\la{nablaC}
\end{equation}
we have an operator that parallel transports the zweibein.

We now proceed to employ this formalism to 
rewrite the Yang-Mills Lagrangian (\ref{ein1})-(\ref{ein4})
in a generally covariant manner. Our computations simplify 
considerably when we observe that for the metric tensor (\ref{met1})
\[
\nabla_\mu  \ce^{}_\nu  + \omega^{\, \, \lambda}_{\mu \,\, 
\nu}\ce^{}_\lambda 
\ = \ 
\partial_\mu \ce^{}_\nu - \Gamma^\lambda_{\mu\nu}
\ce^{}_\lambda + \omega^{\, \, \lambda}_{\mu \,\, \nu}\ce^{}_\lambda 
\ = \ \rho \cdot \partial_\mu ( \frac{ \ce^{}_\nu }{\rho})
\]
and
\[
{\mathcal C}_\mu \ = \ i \bar\ce^\nu \partial_\mu \ce^{}_\nu - 
\frac{i}{4} \partial_\mu \ln \sqrt{g}
\]

We start with the second and third term in (\ref{ein1}), which
we write in
a generally covariant form as follows,
\[
\frac{1}{2} \vec n \cdot \mathrm D^{\hat C}_{\, a} \vec n 
\times \mathrm D^{\hat C}_{\, b} \vec n - n_3
(\partial_a \hat C_b - \partial_b \hat C_a) \
\to \ \frac{1}{2} \vec n \cdot \nabla^{\mathcal C}_\mu 
\vec n \times \nabla^{\mathcal C}_\nu \vec
n - n_3 (\partial_\mu \mathcal C_\nu - \partial_\nu \mathcal C_\mu)
\]
We also write the last contribution in (\ref{ein1}) as  
\[
-2 \rho^2 n_3 H_{ab} = -i \rho^2 n_3 ( e^{}_a e^*_b
- e^{}_b e^*_a) \ \to \  - i \Delta \cdot 
n_3 (\ce^{}_\mu \ce^*_\nu - \ce^{}_\nu
\ce^*_\mu) \ = \ - 2 \Delta \cdot n_3 {\mathcal H}_{\mu\nu}
\]
When we define
\[
{\mathcal F}_{\mu\nu} =
\partial_\mu J_\nu - \partial_\nu J_\mu + 
\frac{1}{2} \, \vec n \cdot \{
\nabla^{\mathcal C}_\mu \vec n \times
\nabla^{\mathcal C}_\nu \vec n - 2 {\hat {\vec z}} [
(\partial_\mu \mathcal C_\nu - \partial_\nu
\mathcal C_\mu) + 2 \Delta \cdot {\mathcal H}_{\mu\nu} ]\}
\]
where ${\hat {\vec z}}$ is a unit vector in the $z$-direction
of the internal space,
we conclude that we can write the entire 
(\ref{ein1}) in the following generally covariant form
\begin{equation}
L^{(1)}_{YM} \ \to \ \mathcal L^{(1)}_{YM} \ = \
\frac{1}{4} \sqrt{g} g^{\mu\nu}g^{\rho\sigma}
{\mathcal F}_{\mu\rho}{\mathcal F}_{\nu\sigma}
\la{L1}
\end{equation}
Similarly we can write (\ref{ein2}) in the following generally 
covariant form,
\begin{equation}
L^{(2)}_{YM} \ \to \ \mathcal L^{(2)}_{YM} \ = \ 
\Delta^2 \cdot \sqrt{g} \, g^{\mu\nu} (J_\mu J_\nu +
\nabla^{\mathcal C}_\mu \vec n \cdot \nabla^{\mathcal C}_\nu 
\vec n)
\la{L2}
\end{equation}

We now proceed to (\ref{ein3}). For this we send our flat space
$U^{}_I(1)$ covariant derivative
of the Euclidean metric Grassmannian zweibein to a
generally covariant form as follows,
\[
(\partial_a + i C_a ) e_b \ \to \ {E_a}^\lambda \{ \delta_{\,
\, \, \lambda}^{\nu}
\nabla^{\mathcal C}_\mu + \omega^{\,\, \nu}_{\mu \, \,
\lambda} \} \ce^{}_\nu   \ = \ {E_a}^\lambda \mathcal D_{\mu \,\,
\lambda}^{\,\, \nu} \ce^{}_\nu
\]
We have here introduced the following twisted covariant derivative
\[
{\mathcal D}_{\mu \,\, \lambda}^{\,\, \nu}
\ = \ \delta_{ \,\,\, \lambda}^{\nu} \nabla^{\mathcal C}_\mu +  
\omega^{\, \, \nu}_{\mu \,\, 
\lambda} 
\]
It extends the action of the twisted
covariant derivative (\ref{nablaC})
to the vector fields $\ce^{}_\nu$.
With this, we can present the entire (\ref{ein3}) in
the following covariant form,
\[
L^{(3)}_{YM} \ \to \ \mathcal L^{(3)}_{YM} 
\]
\begin{equation}
= \ \Delta^2 \cdot \sqrt{g} \cdot g^{\mu\nu} g^{\lambda \eta}
\left\{ ({\bar {\mathcal D}}_{\mu \,\, \lambda}^{\,\, \sigma} \bar{\ce}_\sigma)
({\mathcal D}_{\nu \,\, \eta}^{\,\, \kappa} \ce^{}_\kappa)
+ \frac{1}{2} t^{}_+ ({\bar{\mathcal D}}_{\mu \,\, \lambda}^{\,\, 
\sigma} \bar{\ce}^{}_\sigma) ({\bar{\mathcal D}}_{\nu \,\, \eta}^{\,\, \kappa} 
\bar{\ce}^{}_\kappa) + 
\frac{1}{2} t^{}_- ({\mathcal D}_{\mu \,\, \lambda}^{\,\, 
\sigma} \ce_\sigma) ({\mathcal D}_{\nu \,\, \eta}^{\,\, \kappa} 
\ce^{}_\kappa) \right\}
\la{ein3b}
\end{equation}

Finally, we proceed to (\ref{ein4}). We introduce the Riemann
tensor in terms of the Christofffel symbol (\ref{chris})
\[
R_{\mu \nu \sigma}^{\,\,\,\,\,\,\,\, \,
\,\, \lambda} = \partial_\nu \Gamma^\lambda_{\mu \sigma}
- \partial_\mu \Gamma^{\lambda}_{\nu \sigma} + \Gamma^{\eta}_{\mu \sigma}
\Gamma^\lambda_{\eta \nu} - 
\Gamma^{\eta}_{\nu \sigma}
\Gamma^\lambda_{\eta \mu} 
\]
and the Ricci tensor
\[
R_{\mu\nu} = R_{\nu\mu} = R_{\mu \lambda \nu}^{\,\,\,\,\,\,\,\, \,
\,\, \lambda}
\]
and the Ricci scalar
\[
R = R^{\,\, \, \mu}_\mu
\]
which transforms according to (in $D$ dimensional space)
\begin{equation}
R \ \to \ \tilde R = \phi^{-2} \{ R - 2(D-1) 
g^{\mu\nu} \nabla_\mu \nabla_\nu
\ln \phi - (D-2)(D-1) 
g^{\mu\nu} (\nabla_\mu \ln \phi) (\nabla_\nu \ln \phi) \}
\la{RRscale}
\end{equation}
under the conformal scaling 
\[
g_{\mu\nu} \to \tilde g_{\mu\nu} = \phi^2 g_{\mu\nu}
\] 
of the metric tensor. For the metric tensor (\ref{met1}) this leads
to the identification
\[
\frac{1}{2} ( \partial_\mu \rho )^2  \ \to \
\frac{1}{\Delta^2} \sqrt{g} \, R
\]
This is the covariant interpretation of the first term in (\ref{ein4}).
For the remaining terms in (\ref{ein4}) (except for the surface term)
we get from (\ref{met1})
\[
\frac{3}{8} (1-n_3^2) \rho^4  
-   \frac{3}{8}  \rho^4  \ \to \ \frac{1}{\Delta^2}
\frac{3}{8} (1-n_3^2) \sqrt{g}  -   \frac{1}{\Delta^2}
\frac{3}{8}  \sqrt{g}  
\]
and we conclude that the entire
(\ref{ein4}) can be presented in the following generally
covariant manner,
\begin{equation}
L^{(4)}_{YM} \ \to \ \mathcal L^{(4)}_{YM} \ = \ 
\frac{1}{\Delta^2} \sqrt g \, R \ - \ \frac{3}{8} \frac{1}{\Delta^2} 
\sqrt{g} \ + \
\frac{3}{8} \frac{1}{\Delta^2} \,
(1-n^{2}_3) \sqrt g 
\la{L4}
\end{equation}  
Here the first contribution is the standard Einstein-Hilbert Lagrangian,
the second is the standard (negative) cosmological constant term,
and the third gives a (in general) space-time dependent correction to 
the cosmological constant when $n_3 \not= \pm 1$.

Note that the sign of the Ricci scalar is consistent
with the sign proposed in \cite{haw}, ensuring that the
Euclidean Einstein-Hilbert Lagrangian is bounded from below.

We conclude by summarizing, that when we combine (\ref{L1}), (\ref{L2}), 
(\ref{ein3b}) and (\ref{L4}) we find that in terms of the spin-charge
separated variables the Yang-Mills Lagrangian can be written in the
following generally covariant form
\[
L^{}_{YM} = \mathcal L^{(1)}_{YM} + \mathcal L^{(2)}_{YM} + \mathcal
L^{(3)}_{YM} + \mathcal L^{(4)}_{YM}
\]
Here
\[
\mathcal L^{(1)}_{YM} \ = \
\frac{1}{4} \sqrt{g} g^{\mu\nu}g^{\rho\sigma}
{\mathcal F}_{\mu\rho}{\mathcal F}_{\nu\sigma}
\]
which has the standard form of 
the generally covariant Maxwell Lagrangian,
\[
\mathcal L^{(2)}_{YM} \ = \ 
\Delta^2 \cdot \sqrt{g} \, g^{\mu\nu} (J_\mu J_\nu +
\nabla^{\mathcal C}_\mu \vec n \cdot \nabla^{\mathcal C}_\nu 
\vec n)
\]
is a generally covariant current-current interaction term,
\[
\mathcal L^{(3)}_{YM} 
\ = \ \Delta^2 \cdot \sqrt{g} \cdot g^{\mu\nu} g^{\lambda \eta}
\left\{ ({\bar {\mathcal D}}_{\mu \,\, \lambda}^{\,\, \sigma} \bar{\ce}_\sigma)
({\mathcal D}_{\nu \,\, \eta}^{\,\, \kappa} \ce^{}_\kappa)
+ \frac{1}{2} t^{}_+ ({\bar{\mathcal D}}_{\mu \,\, \lambda}^{\,\, 
\sigma} \bar{\ce}^{}_\sigma) ({\bar{\mathcal D}}_{\nu \,\, \eta}^{\,\, \kappa} 
\bar{\ce}^{}_\kappa) + 
\frac{1}{2} t^{}_- ({\mathcal D}_{\mu \,\, \lambda}^{\,\, 
\sigma} \ce_\sigma) ({\mathcal D}_{\nu \,\, \eta}^{\,\, \kappa} 
\ce^{}_\kappa) \right\}
\]
gives the kinetic term for the Grassmannian $\ce_\mu$ together
with two terms describing its interaction with $\vec n$ where
we recall that these interaction terms can be related to the 
(anti)holomorphic one-forms on the Grassmannian. Finally,
\[
\mathcal L^{(4)}_{YM} \ = \ 
\frac{1}{\Delta^2} \sqrt g \, R \ + \ - \frac{3}{8} \frac{1}{\Delta^2}
\sqrt{g} \ + \
\frac{3}{8} \frac{1}{\Delta^2} \,
(1-n^{2}_3) \sqrt g 
\]
is the Einstein-Hilbert Lagrangian together with a cosmological constant
with a space-time dependent correction that vanishes in the
ground state where $n^{}_3 = \pm 1$.

The final Lagrangian has a manifestly generally covariant form, and
it coincides with the spin-charge separated
Yang-Mills Lagrangian (\ref{ein1})-(\ref{ein4}) when we evaluate
it using the conformally flat metric tensor (\ref{met1}). 
Notice that from the present point of view the nontriviality of
the ground state expectation value in (\ref{rhocond}) becomes
quite natural. When $\rho$ vanishes our conformal space-time
just shrinks away.

\vfill\eject

%
%
\section{Static limit}
%
%

It is often instructive to inspect the static limit of the Lagrangian,
it gives an indication on the ground state properties of the theory.
We reach the static limit when we only retain the spatial derivatives 
and set the time component of the vector field $\vec e$ to zero. 
This sends $\vartheta \to \pi/2$ in (\ref{ebang}), and as a consequence
the electric vector field $\vec p$ vanishes and the only non-vanishing
contribution to the tensor field $H_{ab}$ is
\[
H_{ij} = \frac{1}{2\sqrt{2}} \, \epsilon_{ijk} l_k
\]
where $\vec l$ is the unit vector in the magnetic direction.
Furthermore, from (\ref{kiri1}), (\ref{kiri2}), (\ref{maglim}) 
we get
\[
\partial_i {\hat{\! \mathit C}}_j - \partial_j
{\hat{\! \mathit C}}_i \ = \ - \frac{1}{2\sqrt{2}}\,\,
\vec l \cdot \partial_i \vec l \times
\partial_j \vec l
\]
and
\[
(\partial_i e_j)^2 \ = \ \frac{1}{16} \cdot
\{ ( \, \vec l
+ i \vec m ) \cdot \partial_i \vec k \}^2
\]
With these, we get from (\ref{YMfi}) for the energy density
in the static limit
\begin{eqnarray}
H_{static} = \frac{1}{2}(\partial_i \rho)^2
+ \frac{1}{2}\rho^2  J_i^2 + \frac{1}{8}\rho^2  (\mathrm 
D^{\hat C}_{ \, i} 
\vec n)^2
+ \frac{1}{32} \rho^2 (\partial_i \vec l)^2 \hskip 2.cm \
\la{Hsfi1}
\\
+ \frac{1}{64} \rho^2
\left\{ n^{}_+ \, e^{-2i\psi}  \, 
( \, [\, \vec l + i \vec m]\cdot \partial_i \vec k )^2
+ n^{}_- \, e^{2i\psi} \,( \, [\, \vec l -
i \vec m]\cdot \partial_i \vec k )^2
\right\} \hskip 1.3cm \
\la{Hsfi2}
\\
+ \frac{1}{4}{\mathcal F}_{ij}^2 + \frac{3}{8} (1-n_3^2) \rho^4 
- \frac{3}{8} \rho^4 \hskip 4.3cm \
\la{Hsfi3}
\end{eqnarray}
Here
\begin{equation}
\mathcal F_{ij} =  \partial_i J_j - \partial_j J_i
+ \frac{1}{2} \vec n \cdot \mathrm D^{\hat C}_{ \, i} \vec n \times
\mathrm D^{\hat C}_{ \, j} \vec n + n_3 \frac{1}{2\sqrt{2}} \{ \,
\vec l \cdot \partial_i \vec l \times
\partial_j \vec l - 2 \rho^2 \epsilon_{ijk} l_k \}
\la{curlf2}
\end{equation}
From this we can draw the following conclusions:  

There is an apparent duality between the internal 
vector field $\vec n$ and the space-valued vector field $\vec l$. 
In particular, both are embedded in
(\ref{Hsfi1})-(\ref{Hsfi3}) in a manner that employs 
the version (\ref{0fha}) 
of the $O(3)$ nonlinear
$\sigma$-model. As a consequence {\it 
both } $\vec n$ and $\vec l$ have the potential 
of supporting closed knotted strings as stable solitons. It is suggestive
to interpret $\vec l$ as a ``magnetic'' order parameter, and $\vec n$ as an
``electric'' order parameter in the static limit \cite{omat2}. 

The vector field $\vec n$ takes values in the internal space. Due to
the one-cocycle (\ref{cycle1}) in the Lorentz transformations, it has 
a unique rotation invariant ground state value at large distances
which is given by (\ref{defnasym}). But $\vec l$ is a space valued 
vector field, it transforms as a vector under spatial $SO(3)$ rotations.
Consequently its only conceivable asymptotic ground state value 
at large distances is the spherically symmetric
\[
\vec l \ \buildrel{r \to \infty} \over \longrightarrow 
\ \frac{\vec x}{r}
\]
From this we get the asymptotic behaviour
\[
\vec l \cdot \partial_i \vec l \times
\partial_j \vec l \  \buildrel{r \to \infty} \over \longrightarrow 
\epsilon_{ijk} \frac{x^k}{r^3} \ \sim \ \frac{1}{r^2}
\, \epsilon_{ijk} l_k
\]
for the third contribution to (\ref{curlf2}). Note that this is
reminiscent of a magnetic monopole. 

We combine the last two terms in (\ref{curlf2}) asymptotically 
into
\begin{equation}
\vec l \cdot \partial_i \vec l \times
\partial_j \vec l \  - 2 \rho^2 \epsilon_{ijk} l_k
\ \buildrel{r \to \infty} \over \longrightarrow  \
(\frac{1}{r^2} - 2 \rho^2 ) \epsilon_{ijk} \frac{x^k}{r} 
\la{asyl}
\end{equation}

We now make the following proposals: For a finite energy, 
we can expect that each of the positive definite
terms in the static Hamiltonian are integrable.
This means that asymptotically at large distances, in 
an analytic power expansion in $r$ we can expect 
\[
\rho(r) \ \lesssim \ {\mathcal O} (\frac{1}{r})
\]
In terms of the four dimensional metric tensor (\ref{met1})
this suggests that for finite energy the space should be
compact. But from the present static point of view
we can also argue as follows.
We consider the space-time to have the topology of
$\mathcal M \times \mathbb R^1$ where $\mathcal M$ is 
the three dimensional space manifold and $\mathbb R^1$
is the time. If we define the three-dimensional, spatial
metric tensor by setting
\begin{equation}
g^{(3)}_{ij} = \frac{\rho^4}{\Delta^4} \delta_{ij}
\la{3dg}
\end{equation}
we find that some of the terms in the energy density admit an
independent, {\it three} dimensional geometric interpretation:
From (\ref{RRscale}), the first term in the {\it r.h.s.} of (\ref{Hsfi1})
can be written in terms of the three dimensional Ricci scalar as
\[
 \frac{1}{2} (\partial_i \rho)^2 \ = \
\frac{1}{80} \frac{1}{\Delta^2} \sqrt{ g^{(3)} } R^{(3)}
\]
The second term can be written as
\[
\frac{1}{2} \rho^2 J^2_i \ = \ \frac{1}{2} \sqrt{ g^{(3)} } \,
g^{(3) \, ik} J_i J_k
\] 
Similarly we conclude that each of the terms in (\ref{Hsfi1}) and
(\ref{Hsfi2}) admit a generally covariant interpretation in terms
of the present {\it three} dimensional conformal geometry. 

For the terms in (\ref{Hsfi3})
the present three dimensional geometric interpretation appears to fail. 
But when we demand that the quadratic terms are independently
integrable, since the asymptotic behaviour of $\vec n$ is dictated
by (\ref{defnasym})  we can argue that it 
becomes very natural 
to expect that asymptotically the two terms in (\ref{asyl}) cancel
each other. This suggests that at large distances we have
\[
\rho^2 \ \sim \ \frac{1}{2} \, \frac{1}{ r^2 + \lambda^2}
\]
where $\lambda$ is some parameter.
When we substitute this in (\ref{3dg}) we find that the
spatial part of our space-time becomes asymptotically compactified into the
sphere $\mathbb S^3$. It would be very interesting if this 
proposal could be made more rigorous.

\vfill\eject

%
%
\section{summary}
%
%

We conclude our article with a summary of our results
and a number of remarks on their possible physical
consquences.

We have introduced a novel, {\it complete} field 
decomposition in the Yang-Mills Lagrangian. The decomposition 
implements a separation between the spin 
and the charge in the gauge field. The decomposition
also introduces an internal, compact $U(1)$ interaction. 
A compact $U(1)$ gauge theory is known to exhibit 
confinement in a strong coupling domain which 
is separated from a weakly coupled and deconfined 
domain by a first order phase transition. Since 
the coupling in an abelian theory should increase when
the distance scale decreases, the spin-charge 
separation is not in an apparent conflict with 
the high energy limit of the Yang-Mills theory, 
represented by asymptotically free and massless gauge
bosons.

The spin-charge separated Yang-Mills Lagrangian
describes the interacting dynamics between
a version \cite{fadpap} of the $O(3)$ 
nonlinear $\sigma$-model and a $G(4,2)$ Grassmannian
nonlinear $\sigma$-model, in a conformally 
flat spacetime and in the presence of both the Einstein-Hilbert 
Lagrangian and a negative cosmological constant term. 

The conformal scale of the metric coincides with the
gauge invariant condensate that has been studied previously 
in \cite{kondo}, \cite{zakh}.  Numerical lattice studies indicate that the 
ground state value (\ref{rhocond}) of the condensate is nonvanishing.
From our geometrical point of view this is an expected result:
If the conformal scale vanishes there is no space-time.

The metric properties of the classical Yang-Mills
theory are consistent with the short distance limit 
of a renormalizable higher derivative gravitation theory
with a Lagrangian of the form (\ref{ehacti}). 
It would be truly exciting if at short distance 
Einstein gravity metamorphoses into a 
Yang-Mills theory, as a single renormalizable quantum 
theory of material interactions.

The presence of the higher
derivative term in (\ref{ehacti}) gives rise to a linearly increasing
component in the large distance gravitational interaction. 
From the point of view of the Yang-Mills theory this 
may have some obvious advantages. However, at
distance scales which are well beyond those that should be described by the 
Yang-Mills theory as such, it may become desirable for the $\beta$-function 
of the coupling $\gamma$ in (\ref{ehacti}) to force this coupling
to flow towards $\gamma = 0$. Such a large distance behaviour 
in the quantum theory would then leave the conventional Einstein 
gravity as the sole surviving very long range component 
of the spin-charge separated Yang-Mills theory. 

Note that since $\gamma$ flows towards $\gamma \to \infty$ 
at short distances, the asymptotic condition (\ref{weylc})
dissolves the massless modes of the Einstein gravity from the 
(not too) short distance spectrum. This leaves us with a 
gapped and spin-charge separated Yang-Mills theory that
describes asymptotically free gauge vectors
as distance scale goes to zero.

Our results suggest that due to fluctuations 
in the gauge invariant condensate $\rho$ and the vector field
$\vec n$, at short distances both Newton's constant and the cosmological
constant become variable.

The version of the $O(3)$ nonlinear $\sigma$ model that embeds
the vector field $\vec n$ in the Yang-Mills Lagrangian, 
essentially coincides with (\ref{0fha}). This Lagrangian 
is known to describe knotted strings as stable solitons. 
Our results then support the proposal
\cite{omat1}-\cite{omat3} that such strings 
are present in the spectrum of the $SU(2)$ Yang-Mills theory.

In the absence of the potential term in (\ref{0fha}), the
spectrum of the $O(3)$ model contains two massless Goldstone
bosons. These bosons originate from the asymptotic 
breaking of the global $O(3)$ symmetry, when we select 
the large distance direction for the unit vector $\vec s$. 
In the case of the Yang-Mills theory, these massless Goldstone bosons
are removed by the one-cocycle that breaks the Lorentz
invariance of the pertinent order parameter $\vec n$. The requirement that
the large distance ground state is Lorentz invariant
uniquely fixes the asymptotic direction of the order parameter.

The spin-charge separated Yang-Mills theory
also describes the $G(4,2)$ Grassmannian nonlinear
$\sigma$-model. The $\mathbb S^2 \times \mathbb S^2$ structure
of the $G(4,2)$ manifold has an interpretation in terms of
electric and magnetic variables, the two spheres $\mathbb S^2$
are related to each other by an electric-magnetic duality. In terms of the
corresponding unit vector fields, the Grassmannian contribution to 
the Yang-Mills Lagrangian admits a very transparent realization. 
In particular, in the static (magnetic) limit we are left with 
only one three-component unit vector field $\vec l$, pointing in
the magnetic direction. This vector field is embedded in the Yang-Mills
Lagrangian by a version of the Lagrangian (\ref{0fha}). Consequently
we have an additional duality between two different
embeddings of (\ref{0fha}), described by $\vec n$ and $\vec l$ 
respectively. In particular, this means that the vector field $\vec l$ 
has also the potential of supporting closed knotted
strings as stable solitons.

The interaction between the $O(3)$ $\sigma$-model 
and the Grassmannian $\sigma$ model involves the 
(anti)holomorphic one-forms on the Grassmannian manifold. These
appear in a combination with the $n_\pm$ components of the 
vector field $\vec n$. The presence of the one-cocycle in the 
Lorentz transformation of $\vec n$ implies that at large 
distances these interaction terms are absent.

Finally, we have argued that demanding finiteness of 
energy in the Yang-Mills theory enforces a 
compactification of the space. 
We have proposed that it is natural for the asymptotic 
topology of the space-time to coincide with that of the
manifold $\mathbb S^3 \times \mathbb R^1$.

\vfill\eject

%
%
\section{Acknowledgements}
%
%

The work by L.D.F. has been supported by 
RFBR grant 05-01-00922, CRDF grant RUM-1-2622-ST-04, and the
program "Problems of nonlinear dynamics" of Presidium of Russian 
Academy of Sciences. The work by A.J.N. has been supported by a 
Grant from ANR, by a VR Grant, by a STINT Institutional Grant 
and by a STINT Thunberg Stipend.  L.D.F.
thanks L. Lipatov and A. Slavnov for discussions. 
A.J.N. thanks M. Chernodub, U. Danielsson, U. Lindstr\"om, M. Volkov 
and K. Zarembo
for discussions, M. Niedermaier, M. Paranjape and A. Tseytlin for 
communications, and S. Slizovskiy for comments. 
A.J.N. also thanks the
Yukawa Institute at the Kyoto University, the Department of
Physics at Tokyo University and the Asian Pasific Center
for Theoretical Physics for hospitality during the early part
of this work.

\vfill\eject

%
%
%
%

\end{document}